\pdfoutput=1 % only if pdf/png/jpg images are used
\documentclass[number,sort&compress,5p]{elsarticle}
\usepackage{color}

\usepackage[above]{placeins}

\usepackage[labelfont=bf]{caption}

\usepackage{
	amsmath,
	amssymb,
	booktabs,
	graphicx,
	lineno,
	mathptmx,
	subcaption,
	siunitx,	
	times,
	url,
	wasysym
}

\usepackage[T1]{fontenc}
\usepackage[utf8]{inputenc}

\newcommand{\kapton}{Kapton\textsuperscript{\textregistered}\,}
\modulolinenumbers[5]
\begin{document}

\begin{frontmatter}

\title{Feasibility studies for a wireless 60 GHz tracking detector readout}
%\subtitle{- Draft -}
%\author{S. Dittmeier, A. Schöning, H. K. Soltveit, and D. Wiedner}
\author[physi]{S.~Dittmeier\corref{cor1}}
\ead{dittmeier@physi.uni-heidelberg.de}
\author[physi]{A.~Schöning}
\author[physi]{H.K.~Soltveit}
\author[physi]{D.~Wiedner}

%\small \textit{ Physikalisches Institut, Universität Heidelberg}} %\\
\cortext[cor1]{Corresponding author}
\address[physi]{Physikalisches Institut der Universität Heidelberg, Im Neuenheimer Feld 226, 69120 Heidelberg, Germany}

% !TeX root = ../main.tex
%\section{Introduction}\label{sec:int}

\begin{abstract}

The amount of data produced by highly granular silicon tracking detectors in
high energy physics experiments poses a major challenge to readout systems.
At high collision rates, e.g. at LHC experiments, only a small fraction of
data can be read out with currently used technologies.
To cope with the requirements of future or upgraded experiments
new data transfer techniques are required which offer high data rates at low
power and low material budget.

Wireless technologies operating in the \SI{60}{GHz} band or at higher
frequencies offer high data rates and are thus a promising upcoming
alternative to conventional data transmission via electrical cables or optical
fibers.
Using wireless technology, the amount of cables and connectors in detectors
can be significantly reduced. 
Tracking detectors profit most from a reduced material budget as
fewer secondary particle interactions (multiple Coulomb scattering,
energy loss, etc.) improve the tracking performance in general.

We present feasibility studies regarding the integration of the wireless
technology at \SI{60}{GHz} into a silicon tracking detector. 
We use  
spare silicon strip modules of the ATLAS experiment as test samples 
which are measured to be opaque in the \SI{60}{GHz} range. 
The reduction of cross talk between links and the attenuation of
reflections is studied. 
An estimate of the maximum achievable link density is given.
It is shown that wireless links can be placed as close as
\SI{2}{cm} next to each other for a layer distance of \SI{10}{cm} 
by exploiting one or several of the following measures: highly directive
antennas, absorbers like graphite foam, linear polarization and 
frequency channeling.
Combining these measures, a data rate area density of up to \SI{11}{Tb/(s \cdot m^2)} seems feasible.
In addition, two types of silicon sensors are tested under mm-wave irradiation 
in order to determine the influence of \SI{60}{GHz} data transmission on the detector performance: 
an ATLAS silicon strip sensor module and an HV-MAPS prototype for the Mu3e experiment. 
No deterioration of the performance of both prototypes is observed.

\end{abstract}

\end{frontmatter}

%\maketitle
% !TeX root = ../main.tex
\section{Introduction}\label{sec:int}

Today's and future high energy particle physics experiments %, especially at proton-proton colliders like the LHC, 
have to face high event rates %and pileup 
in order to improve the \linebreak sensitivity for Standard Model precision measurements
and to increase the discovery potential for physics beyond the \linebreak Standard Model.
Detectors with ever-increasing granularity \linebreak have to be used 
in order to enhance sensitivity limits due to spatial and momentum resolution.
%resulting in increasing data rates that have to be transmitted.
Constraints on space, material budget, power consumption and radiation
hardness are \linebreak nowadays the main limitation for the detector construction and 
the data readout bandwidth.
Thus, there is an increased demand for new readout techniques that allow data transfer at extremely high rates. 

Today, wired electrical and optical readout systems are used in particle
detectors at colliders.
Within the last decades, \linebreak wireless data transmission has evolved significantly
and data rates are becoming comparable with wired data links.\linebreak
Nonetheless, presently used wireless systems like WIFI or LTE are not suitable for
particle detectors because of the limited data throughput and the large antennas. 
But at higher carrier frequencies a wireless detector readout seems feasible.
A large bandwidth of \SI{9}{GHz} provided in the \SI{60}{GHz} band allows data rates of several \si{Gb/s} even with simple modulation schemes.
Due to the short wavelength of $\lambda \approx \SI{5}{mm}$ antennas have a very small form factor.
%Thus, the chip integration level can be very high.

Commercial chips for the \SI{60}{GHz} frequency band are available on the market, 
but none of them is foreseen to be used in particle physics experiments 
as the following requirements are usually not met: 
%for particle physics applications
the transceiver chip has to be radiation hard; 
it must be operated at low power
and should provide a high bandwidth at the same time.
%However, because of the stability of the detector environment the bandwidth can be fully exploited which reduces the complexity of the system.
%At the University of Heidelberg, 
For that reason a new \SI{60}{GHz} transceiver ASIC for particle physics applications 
is currently designed \cite{Soltveit:IEEE2013, Soltveit:Wireless}. % with the goal to
%fulfill above listed general requirements of particle physics experiments. 
%The final specifications are yet under development.

\begin{figure}[tbp]
\centering
\includegraphics[width=.85\columnwidth]{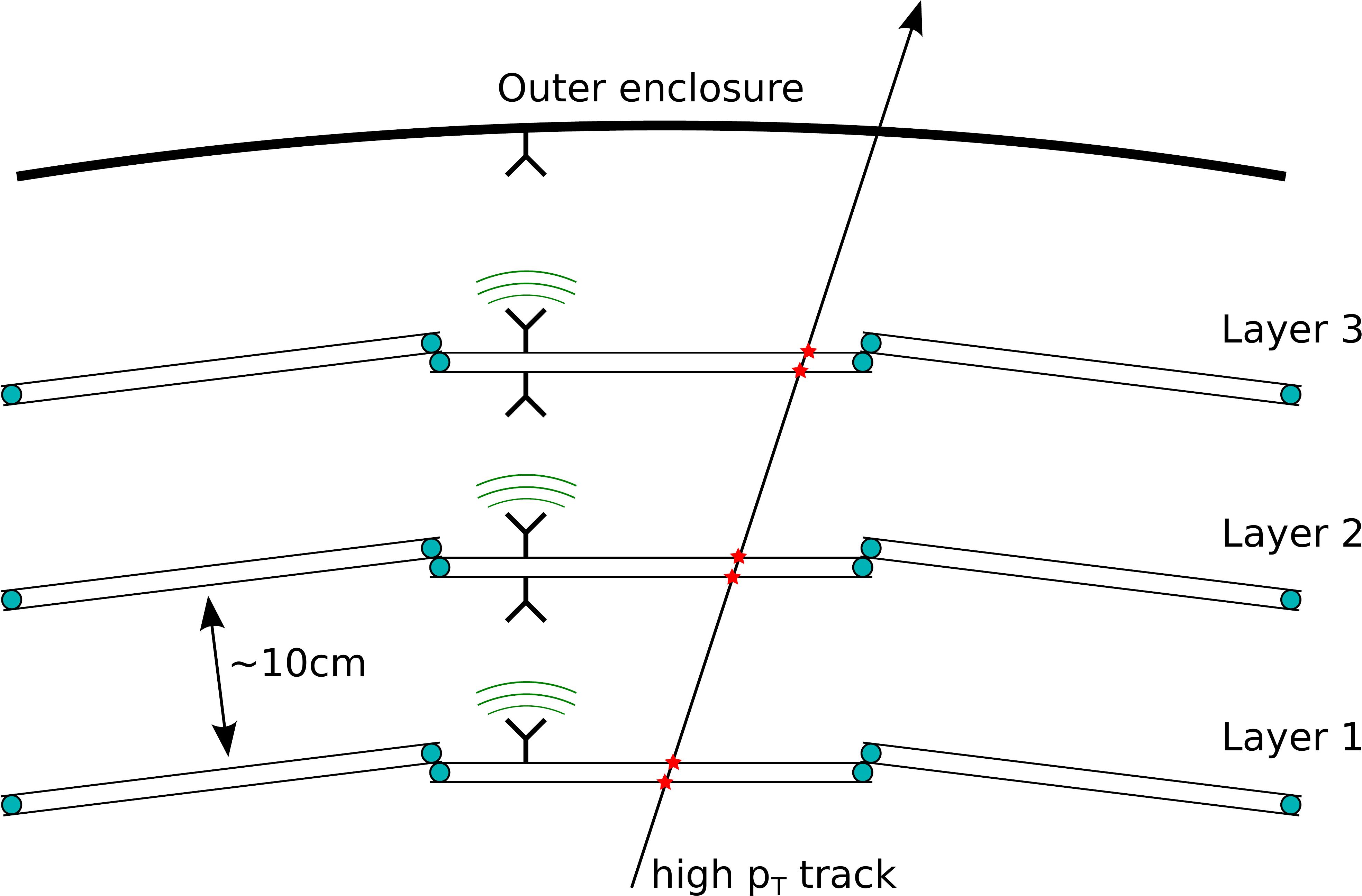}
\caption{Conceptual sketch of a wireless radial readout of a cylindrical tracking detector, adapted from \cite{Brenner:wireless}.}
\label{fig:intro_layers}
\end{figure}

Wireless readout of a tracking detector for a fast track-\linebreak trigger application
was proposed in \cite{Brenner:wireless}.
The authors describe how wireless readout can be exploited to transmit 
hit information between several highly granular silicon tracking \linebreak layers to
enable a fast trigger decision.
The readout scheme, \linebreak depicted in Figure~\ref{fig:intro_layers}, assumes a
radial data transfer from the \linebreak inner detector layers to the outside, thus  
facilitating the \linebreak implementation of track finding algorithms in on-detector logic. % that search for correlations
%between different layers in small roads.

In this paper we present feasibility studies regarding the \linebreak integration of the
\SI{60}{GHz} wireless technology in a silicon tracking detector. 
Several aspects relevant for the implementation of \SI{60}{GHz} links are
studied: transmission losses, interference effects, absorbing materials and the influence of the antenna design. 
In particular, we study how wireless signals can be \linebreak directed with low
material horn antennas and how unwanted \linebreak reflections from detector modules can be attenuated.
From these studies we estimate the maximum density of 
wireless links that can be operated in parallel between detector layers.
In addition, the influence of \SI{60}{GHz} waves on a silicon pixel sensor prototype and on a silicon strip detector module is
tested and it is found that the detector performance is not degraded.

% !TeX root = ../main.tex
\section{60\,GHz transmission and reflection tests}\label{sec:trans}

%GENERAL QUESTION: ARE THE TRANSMISSION MEASUREMENTS CORRECTLY DONE?
%DIFFRACTION EFFECTS SHOULD BE AVOIDED.

\begin{figure}[b]
\centering
\includegraphics[width=.9\columnwidth]{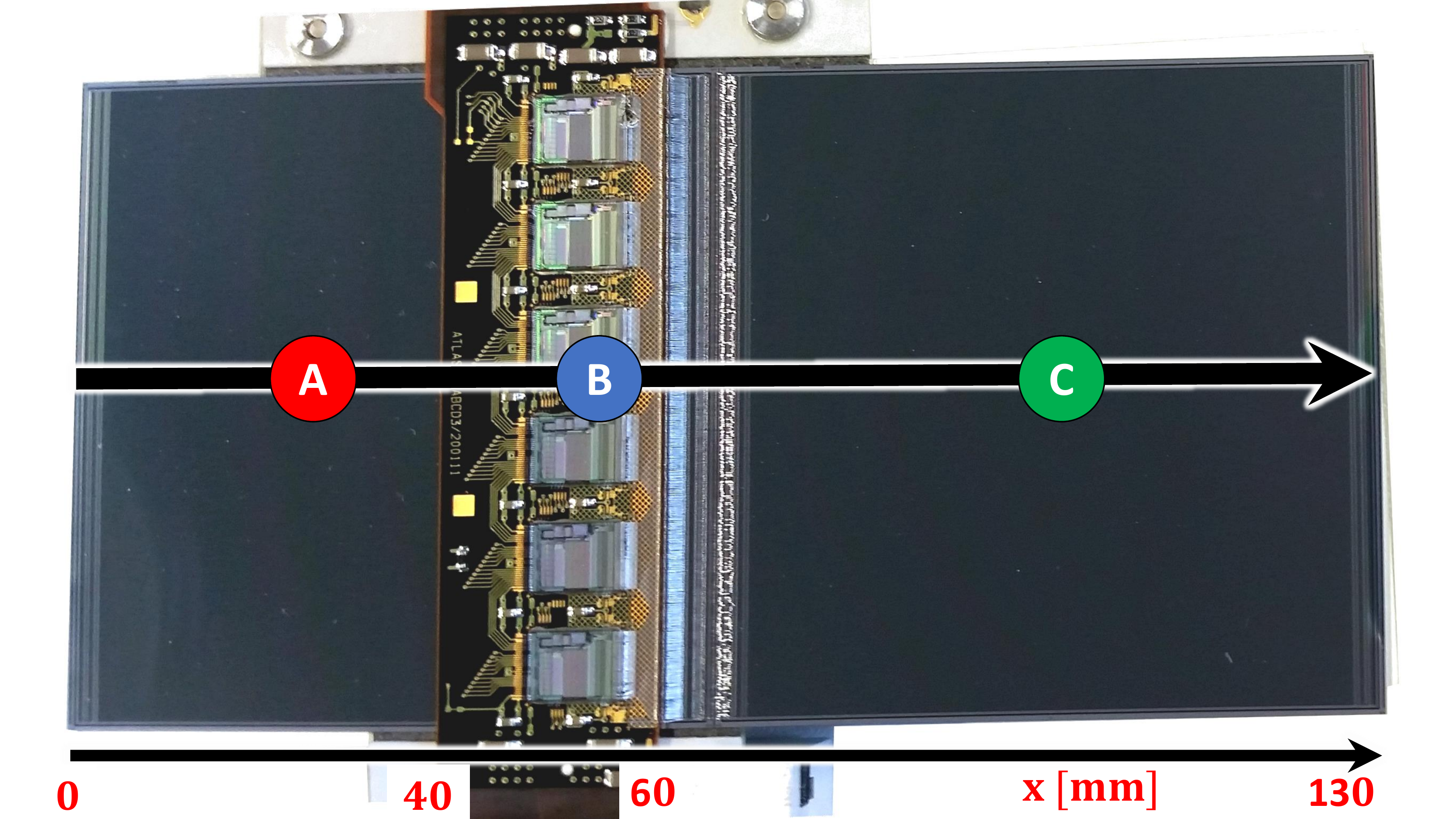}%barrel_scale_hell}
\caption{The ATLAS SCT barrel module \cite{ATLAS:SCT_ups} under test.
Positions for frequency scans are denoted by (A), (B) and (C). A position scan is performed along the black arrow.}
\label{fig:trans_pic_barrel}
\end{figure}

%\begin{figure}[h]
%\includegraphics[width=\columnwidth]{figures/trans_setup}
%\caption{test}
%\label{fig:trans_setup}
%\end{figure}
In order to maximize the data throughput of a wireless readout system, as
depicted in Figure~\ref{fig:intro_layers}, links have to be packed densely and the maximum possible bandwidth should be fully exploited by every single link. 
The main challenge is to avoid cross talk between parallel and subsequent, chained links.
%In the following we tested if a wireless signal can penetrate silicon detector modules which would induce crosstalk between different layers.
The latter is granted as radio signals in the \SI{60}{GHz} band do not pass detector layers with metal layers implemented. 
A first study of transmission of mm-waves through an ATLAS SCT module \cite{ATLAS:SCT_ups}, see Figure~\ref{fig:trans_pic_barrel}, 
showed that mm-waves cannot penetrate tracking detector modules  \cite{Brenner:wireless}. 
We repeat this measurement with increased sensitivity and also with a different type of silicon detector modules.

\subsection{Tests with an ATLAS SCT barrel module}

A spare silicon strip detector module from the ATLAS \linebreak barrel detector, depicted in Figure~\ref{fig:trans_pic_barrel}, is mounted on a 2D \linebreak movable stage and placed in line of sight (LOS) of two horn antennas for transmission and reception.
The module is irradiated with linearly polarized waves in the range from \SIrange{57.3}{61.3}{GHz}~\footnote{If not stated otherwise, all of the following tests are done with the HMC6000/6001 transmitter and receiver chips by Hittite \cite{Hittite:dataTx}.}. 
The intensity transmitted through the module is measured with a spectrum analyzer in the radio frequency band without down conversion. 
The transmitted intensity is normalized to the intensity without module in-between. %, yielding the loss of transmitted intensity through the module. 
The setup is able to resolve transmission losses down to \SI{-55}{dB} at a \linebreak minimum power of about \SI{-90}{dBm} over the frequency range mentioned above. 
To avoid distortions of the measurement by accidental reflections or diffraction aluminium plates and graph\-ite foam is used as shielding. 
%To focus the wireless beam, foam collimators with an aperture of \SI{1}{cm} diameter were placed in front of both transmitting and receiving antennas.

\begin{figure}[tb]
\centering
\includegraphics[width=.95\columnwidth]{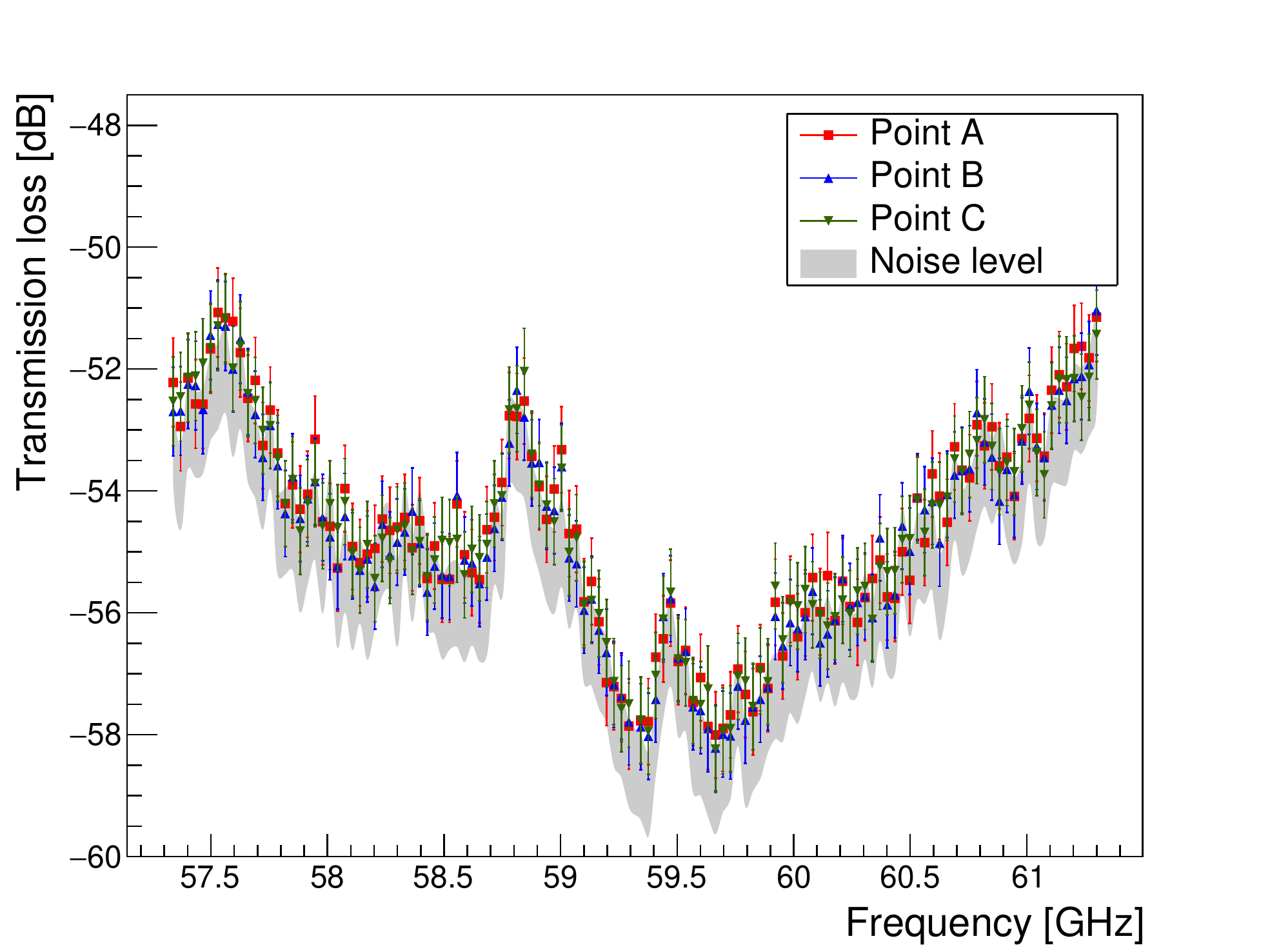}%barrel_spec}
\caption{Transmission loss through the ATLAS SCT barrel module as function of the frequency at positions A, B and C (see Figure~\ref{fig:trans_pic_barrel}) and the noise limited sensitivity of the spectrum analyzer. The uncertainties of \SI{1}{dB} are due to intensity variations observed with the spectrum analyzer in the power range of \SI{-90}{dBm}.}
\label{fig:trans_barrel_spectrum}
\end{figure}

\begin{figure}[tb]
\centering
\includegraphics[width=.95\columnwidth]{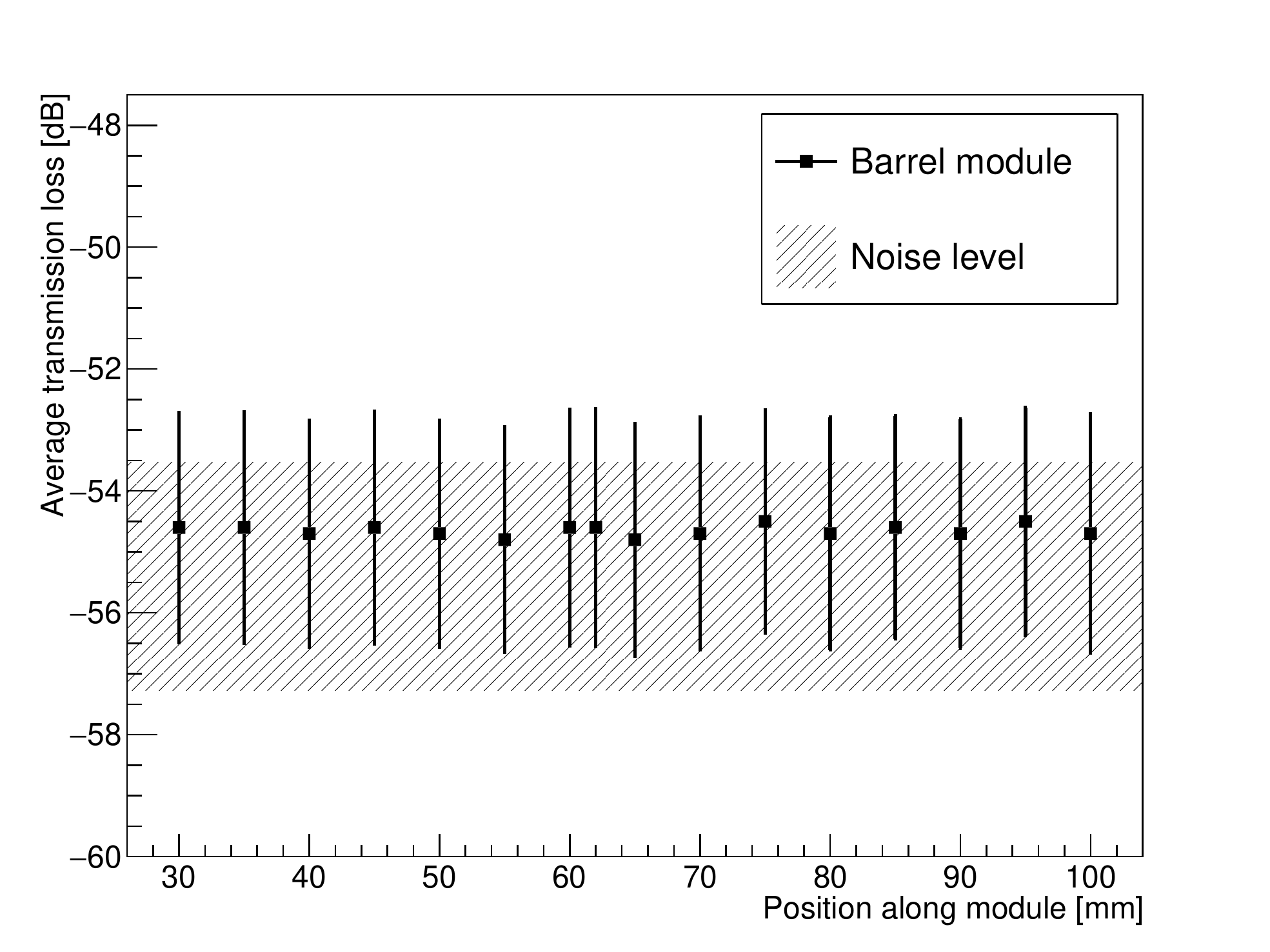}%barrel_avg}
\caption{Transmission loss of the barrel module averaged over the frequency band for a position scan (along the arrow in Figure~\ref{fig:trans_pic_barrel}) and the noise limited sensitivity of the spectrum analyzer. The uncertainties represent the RMS of the average measurements.}
\label{fig:trans_barrel}
\end{figure}

Spectra of the transmission loss through the barrel module at positions A, B and C are shown in Figure~\ref{fig:trans_barrel_spectrum} together with the noise limited sensitivity of the spectrum analyzer. 
The \linebreak region of most interest is region B, as a \SI{60}{GHz} transceiver would have to be placed on the readout electronics hybrid.\linebreak 
Especially with highly directive antennas, most of the \linebreak wireless signal intensity would be focused within this region. 
% For comparison the noise limited sensitivity of the spectrum analyzer is shown. 
% There is no significant deviation of the measured signal from the noise level visible. 
No transmitted signal can be measured at the positions under test, corresponding to a transmission loss of more than \SI{-50}{dB} over the entire frequency range.

Figure~\ref{fig:trans_barrel} shows the transmission loss averaged over the \linebreak chosen frequency band at various positions along the module. \linebreak
Again, no transmission is observed independent of the polarization of the radio signal.
%Close to the edges signals are detected to to diffraction (THIS IS NO TRANSMISSION).
%Instead, the full signal is reflected. 
%This transmission loss appearas at all frequencies and both polarization states.

\subsection{Tests with an ATLAS SCT endcap module}

\begin{figure}[bp]
\centering
\includegraphics[width=\columnwidth]{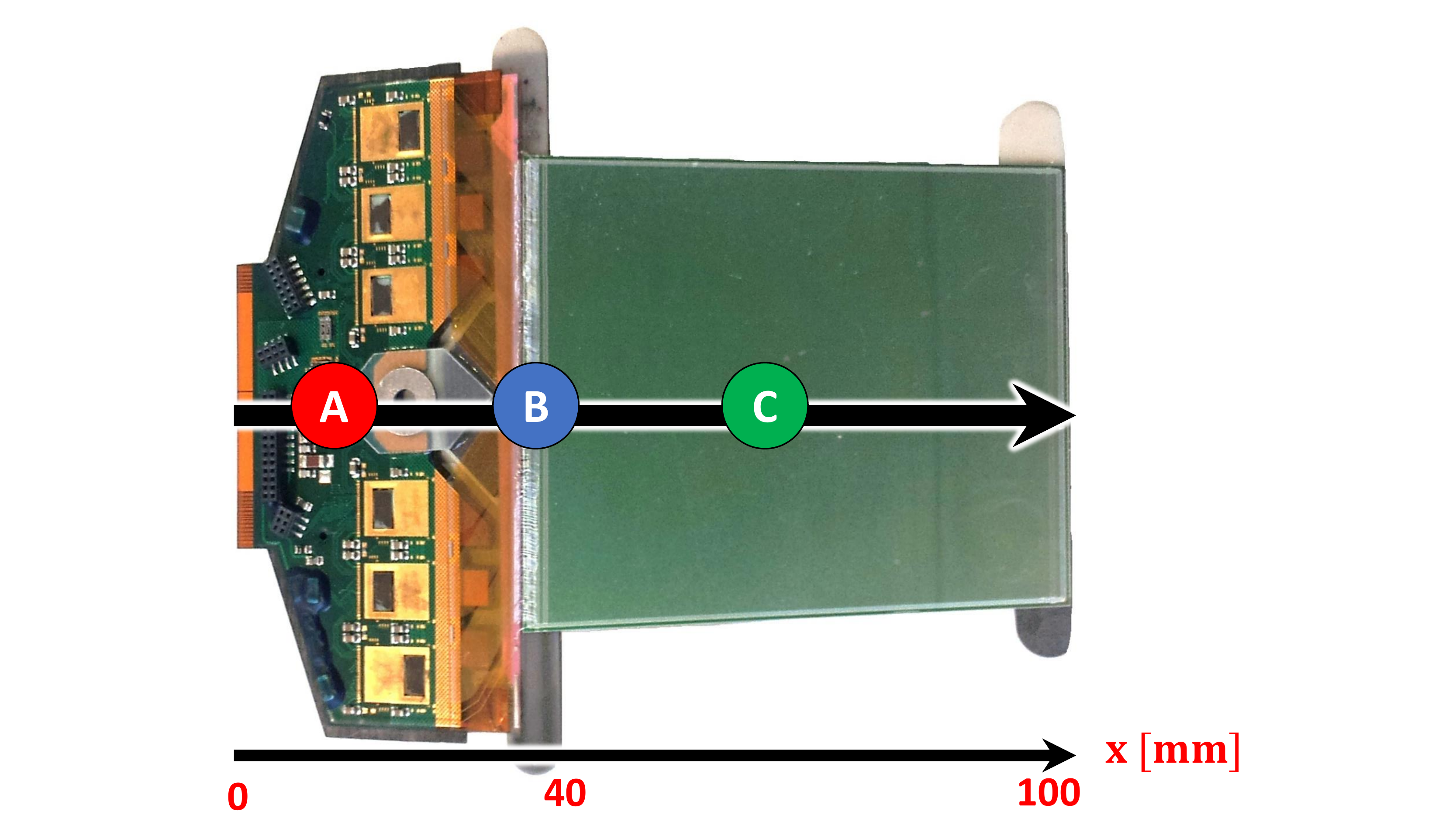}%_scale} angle = 90, origin = c,
\caption{The ATLAS SCT endcap module \cite{ATLAS:SCT_fre} under test}
\label{fig:trans_pic_endcap}
\end{figure}

\begin{figure}[tb]
\centering
\includegraphics[width=.95\columnwidth]{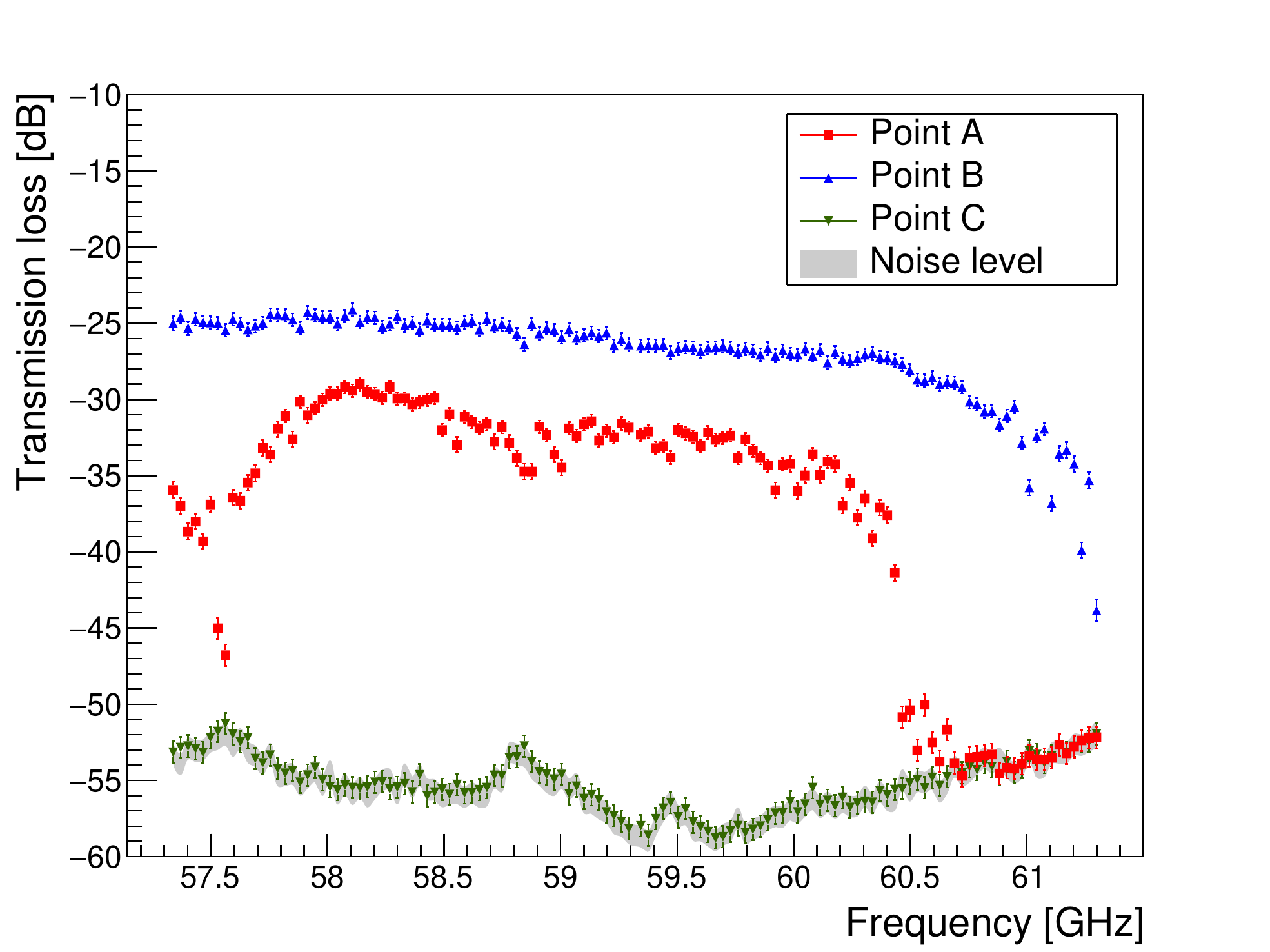}%endcap_spec_2}
\caption{Transmission loss spectra of the endcap module at positions A, B and C (see Figure~\ref{fig:trans_pic_endcap}). The uncertainties are again due to intensity variations observed in the spectrum analyzer.}
\label{fig:trans_endcap_spectrum}
\end{figure}

The measurement is repeated with a spare ATLAS SCT endcap module \cite{ATLAS:SCT_fre}, see Figure~\ref{fig:trans_pic_endcap}. 
%Because of its smaller size diffraction effects at the edges are more prominent.
%Therefore, a reliable position scan could only be performed in the central region of the module.
A position scan is performed only in the central region to avoid diffraction at the edges.\linebreak
Again, the most interesting region for placing a \SI{60}{GHz} \linebreak transceiver is on the readout electronics hybrid, around position A in Figure~\ref{fig:trans_pic_endcap}.
Figure~\ref{fig:trans_endcap_spectrum} shows the transmission loss \linebreak spectra for
positions A, B and C on the module, compared to the measurement's sensitivity level. %noise level of the spectrum analyzer. 
As expected, large variations depending on the frequency are observed in the readout electronics region due to the assembly hole and the gap between hybrid and flex print.

%Only near the central region of the silicon strips (C) no transmission was observed. 
%At the positions A and B, a transmission loss of about \SIrange{-24}{-40}{dB} was
%measured, which is 

\begin{figure}[tb]
\centering
\includegraphics[width=.95\columnwidth]{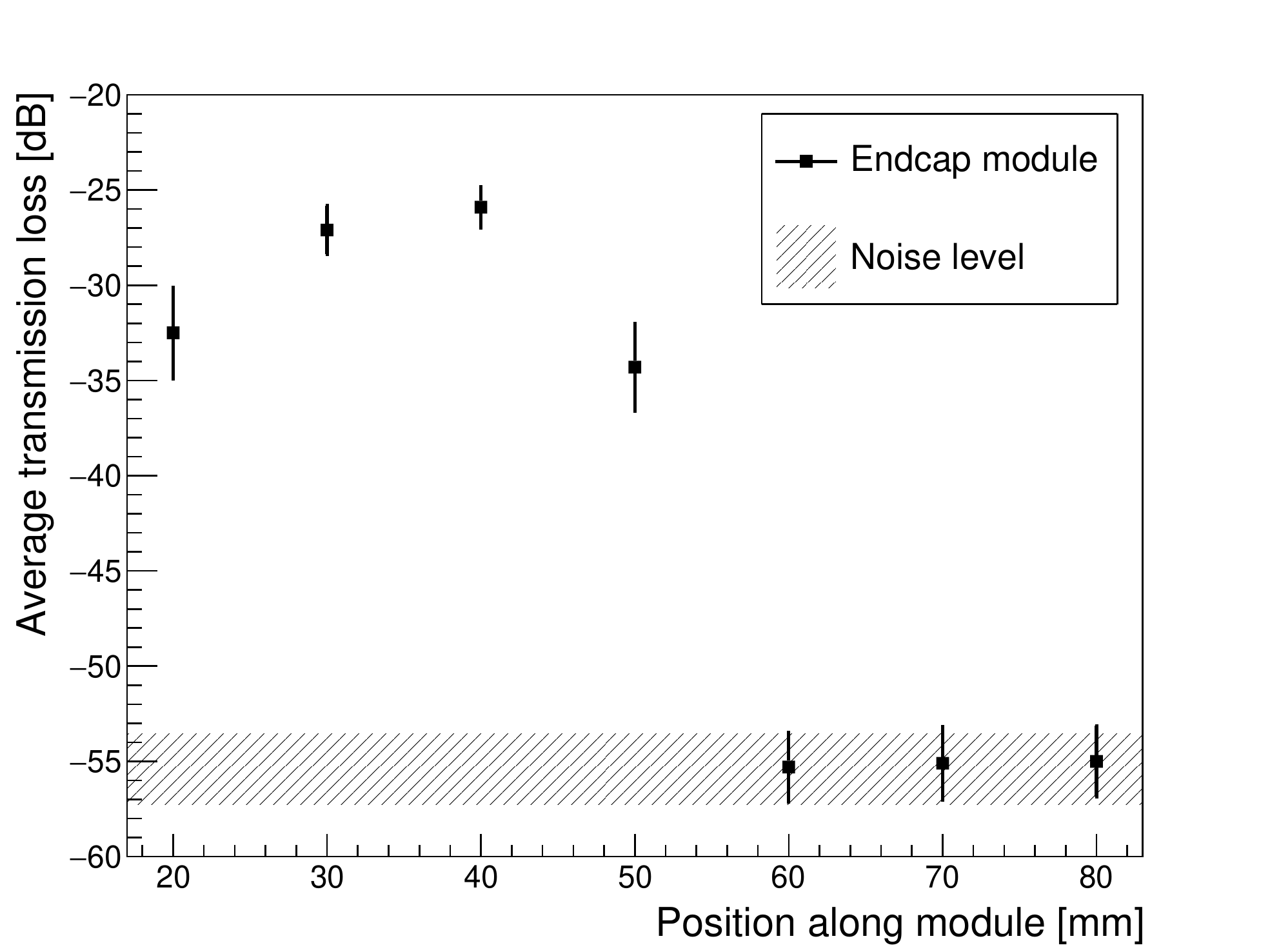}%endcap_avg}
\caption{Transmission loss of the endcap module averaged over the frequency
  band for a position scan (along the arrow in Figure~\ref{fig:trans_pic_endcap}).}
\label{fig:trans_endcap}
\end{figure}

The frequency averaged transmission loss as function of the position on the module, see Figure~\ref{fig:trans_endcap}, is \SIrange{-20}{-40}{dB} in the electronics region for the reasons mentioned above, while 
%Due to the smaller size of the endcap module diffraction effects are significantly larger than for the SCT barrel module.
the silicon strip region is opaque for the mm-waves due to metalization.
%We observed a transmitted signal with less transmission loss of 
%This is due to the gap between the hybrid and the flex print as well as to the hole for the assembly screw.

%\begin{figure}[h]
%\centering
%\includegraphics[width=.5\columnwidth]{figures/reflection}
%\caption{Measurement of intensity reflected under angle of incidence.}
%\label{fig:trans_ref}
%\end{figure}

\begin{figure}[tb]
\centering
\includegraphics[width=.95\columnwidth]{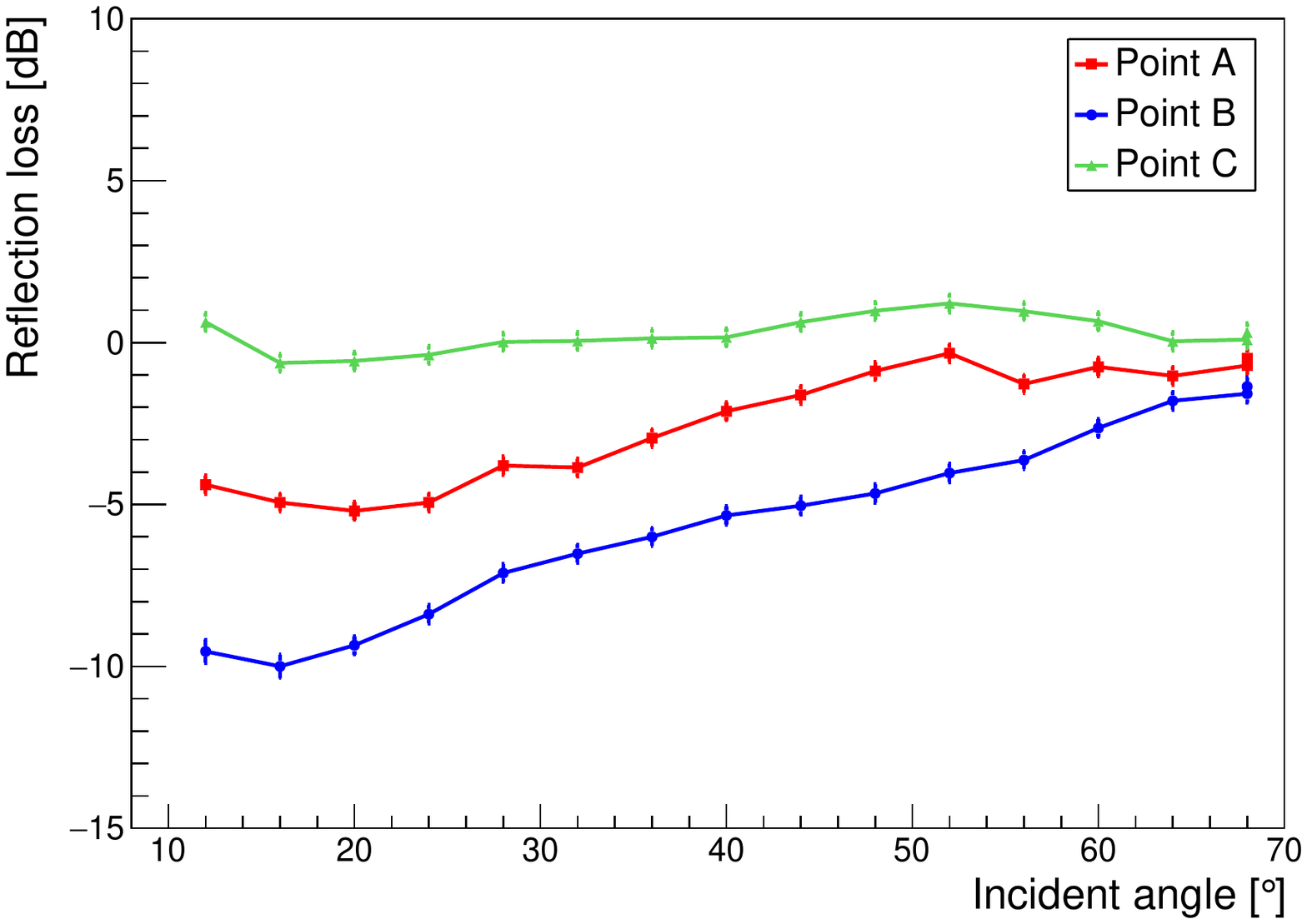}
\caption{Reflection loss of the endcap module at \mbox{$f_{RF} = \SI{59.23}{GHz}$} at positions A, B and C (see Figure~\ref{fig:trans_pic_endcap}) as a function of the angle of incidence. The uncertainties are \SIrange{0.3}{0.5}{dB} due to intensity variations in the power range of \SI{-30}{dBm} to \SI{-40}{dBm}, correspondingly. }
\label{fig:trans_endcap_ref}
\end{figure}

%In general, there is no transmission through standard silicon detectors due to metal layers, 
%which lead to large reflections in the \SI{60}{GHz} band. 
We also measure the reflection for the endcap module \linebreak normalized to the
reflection at a highly reflective aluminum plate %serving as ideal reflector 
for reference. % yielding the reflection loss.
This measurement is performed at a fixed frequency of \mbox{$f_{RF} = \SI{59.23}{GHz}$} with waves polarized perpendicularly to the incident plane.
The intensity of the signal is measured under the assumption that the incident and reflected angle are identical. 
As can be seen in Figure~\ref{fig:trans_endcap_ref}, 
the strip sensor \mbox{(point C)} reflects the signal without any significant losses. 
The readout electronics hybrid (point A) and the bond wire region (point B) of the module show
some reflection losses. 
This could be due to absorption or scattering as the module's surface is not flat in this region. 

Small size components like capacitors, chips, etc. lead to scattering of \SI{60}{GHz} waves. 
Such effects are difficult to \linebreak calculate or simulate as 
standard ray tracing programs fail to describe the data without implementing a detailed 
model for diffraction~\cite{Hugle:Bsc}.

Although pure silicon is not reflective, silicon detectors are highly reflective in the \SI{60}{GHz} band due to metalization and do not allow for any signal transmission through the sensor.
%Other parts belonging to the readout electronics, where a wireless transceiver chip and antenna would be placed, are partially transmissible, but not to a critical level.
Also parts belonging to the readout electronics, where a wireless transceiver chip and antenna would be placed, are in general reflective. % and not transmissible.
Depending on the specific application, these reflections might induce cross talk. Absorbing materials could be used as shields to attenuate reflections.
Gaps and not fully metalized layers, like the flex print of the endcap module, lead to transmission.
Absorbing materials could be used to cover gaps and reduce inadvertent transmission.
The reduction of cross talk by means of absorbing materials is further discussed in section~\ref{sec:absorb_foam}.

%Detailed ray tracing simulations using detector module models can be benefical to optimize the placement of wireless links for a readout system. 
%Especially reflections from the readout electronics of a module are not trivial to describe. 
%Holes and gaps between modules could be closed using an absorber like graphite foam, which was shown in \cite{Dittmeier:WIT2014} to have excellent properties of attenuating reflections.

% !TeX root = ../main.tex
\section{Reduction of cross talk}\label{sec:absorb}

A tracking detector represents a highly stable environment for a wireless communication system. 
Once it is installed, all wireless links are stationary.
Therefore, a wireless readout \linebreak system can be designed and optimized in such a way that interference effects between different links are minimized.
%For instance, the fading margin of the transceiver can be chosen as small as \SI{3}{dB}. %, allowing to cut off non-line-of-sight signals in the receiver.
%However, this requires a very detailed model and study of the link environment.
%The reflectivity of silicon detector modules is measured to be high, which can induce cross talk between parallel links.
%Ray tracing simulation studies with a simple geometrical model indicate that reflections can be an issue assuming a high link density \cite{Hugle:Bsc}.
\linebreak However, assuming an extremely high link density, cross talk can be an issue. 
There are several measures which can also be combined to reduce cross talk, if necessary.
First of all, one can use directive antennas with high gain and small beamwidth. 
By increasing the gain of the antenna the transmitted power 
can be decreased while keeping the signal over noise ratio (S/N) in the receiver constant. 
% Thus, signals picked up by adjacent receivers are reduced due to the lower overall signal amplitude and due to the smaller beamwidth.
Secondly, one can exploit linear polarization. 
By using orthogonal polarization states cross talk between \linebreak adjacent links can
be largely reduced.
%In addition to the gain and beamwidth, the degree of polarisation is another property of the antenna.
Thirdly, reflections can be attenuated by means of absorbing materials. 
Low mass \linebreak materials with high absorbance might be preferably used in \linebreak order not
to increase the material budget.
Fourthly, the \linebreak frequency band can be divided in different channels, using a tunable VCO with sharp filters or low bandwidth antennas. 

Studies on cross talk suppression methods are described in detail in \cite{Dittmeier:Msc}.
Here a summary of these results for the first three options is given.
Frequency channeling is discussed in \linebreak section~\ref{sec:links:ch} in the context of bit
error rate measurements.

\subsection{Directive antennas}

Requirements on antennas are highly application specific.
%The requirements posed on the antenna are highly dependent on the specific application.
Flat patch antennas have a small form factor and can offer high gain and directivity if several patches are combined and interfere constructively.  
However, their bandwidth is typically limited to less than \SI{1}{GHz} \cite{Pelikan:WIT2014}. %and process variations can introduce shifts of the resonance frequency 
Other antenna types as slot antennas \cite{Slot} or Vivaldi antennas \cite{Vivaldi} offer higher bandwidths. 
Due to the small form factor, it is also possible to integrate antennas on-chip. 
But because of losses in the silicon substrate, these antennas usually have a reduced radiation efficiency \cite{Huang}.

%Requirements on the antenna depend highly on the application. 
%To be compatible with our chip an antenna with a high bandwidth is necessary to reach the full target data rate.
%Flat antennas can be produced rather easily on PCB.
%Patch antenna prototypes, simulated, produced and tested (cite Pelikan WIT)
%Directivity can be obtained, but small bandwidth.
%Other flat antennas are an option.
%Or horn antennas made from aluminized Kapton foil.
%Nice properties, high gain, large bandwidth.
%Light material.
%But 3D volume. 
%Mechanical damage can diminish quality of data transfer. 

\begin{figure}[tbp]
\centering
\includegraphics[width=.85\columnwidth]{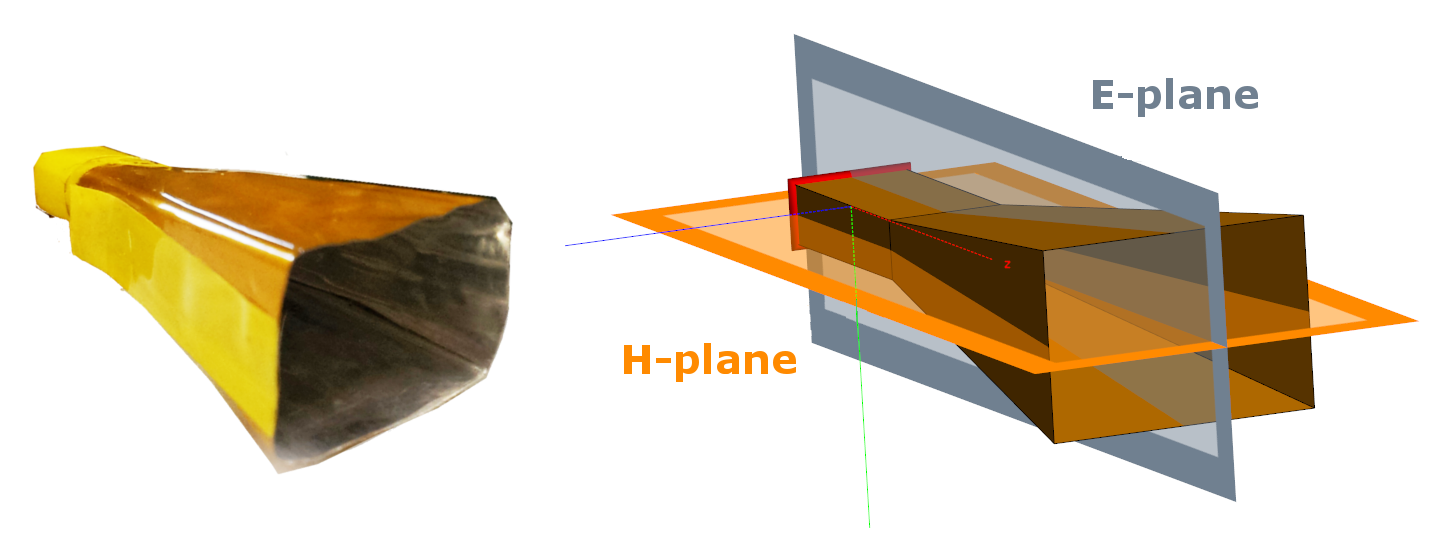}%antenna_direct_hplane}
\caption{Left: photograph of a \SI{3.5}{cm} long horn antenna made from an aluminum and \kapton foil laminate. Right: the H-plane is spanned by the long edge and the direction of emittance (orange plane); the E-plane is spanned by the short edge and the direction of emittance (grey plane).}
\label{fig:antenna_horn}
\end{figure}

For high bandwidth applications horn antennas are a good alternative which, however, 
require some space.
To be usable in the active volume of a tracking detector, the antenna should
consist of as little material as possible.
As an alternative to commercial brass antennas
we produced horn antennas from a \SI{25}{\micron} aluminium and
\SI{25}{\micron} \kapton foil laminate, see \linebreak Figure~\ref{fig:antenna_horn}. 
These light weight antennas serve as a first demonstrator and are used
in the following measurements.
In principle, the aluminium could be even thinner: 
a \SI{25}{\micron} \kapton foil with a \SI{50}{nm} thin aluminium layer is tested to be fully \linebreak reflective at \SI{60}{GHz}. 
Thus, horn antennas could be produced with extremely small material budget \cite{Dittmeier:Msc}. 

\begin{figure}[tbp]
\includegraphics[width=\columnwidth]{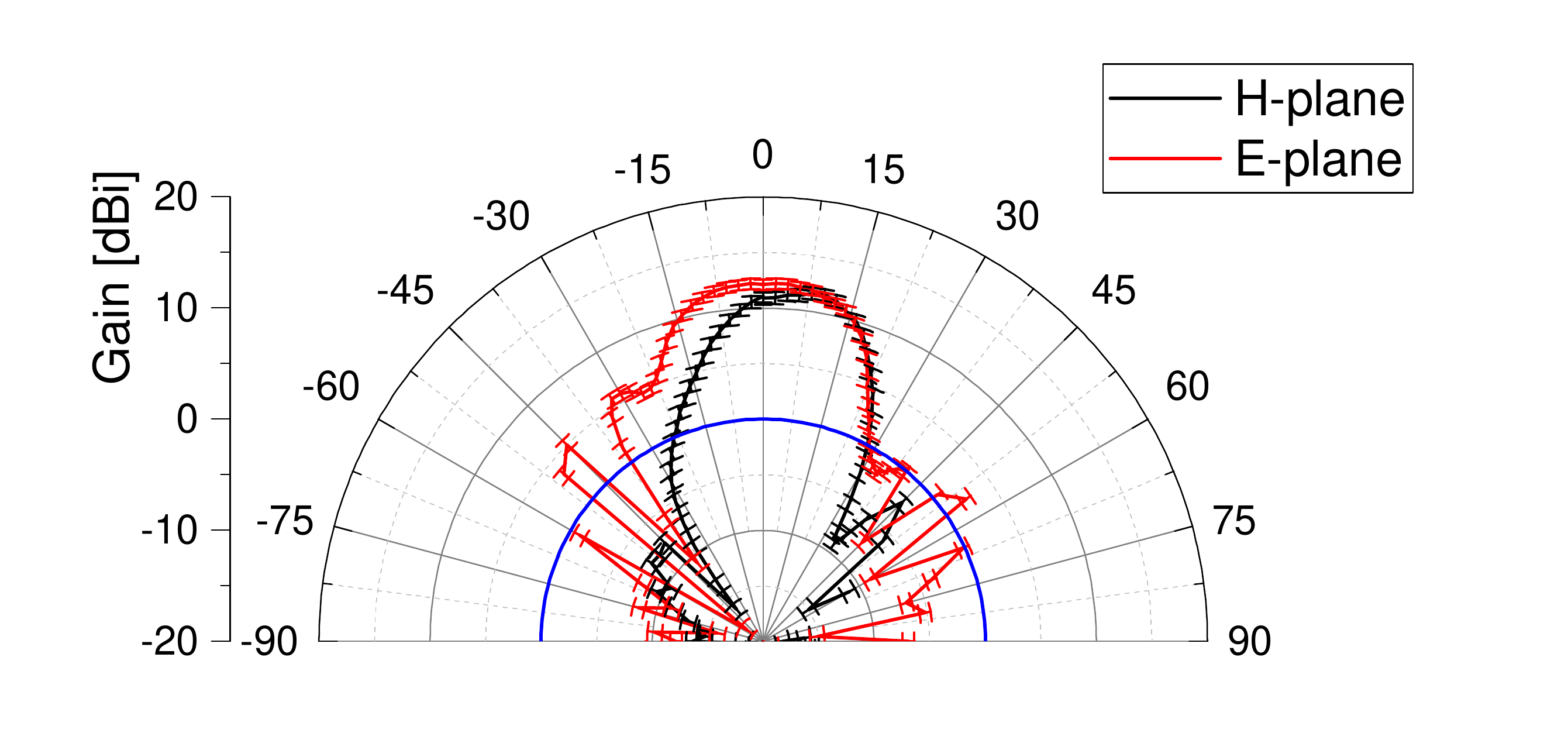}%antenna_direct_hplane}
\caption{Measured polar radiation pattern in the E- and H-plane of 
a tested Al-\kapton horn antenna at $f = \SI{60.85}{GHz}$. 
The blue line represents an isotropic emitter with $G = \SI{0}{dBi}$.}
\label{fig:antenna_direct}
\end{figure}

\begin{figure}[tbp]
\hspace{-2em}
\includegraphics[width=1.1\columnwidth]{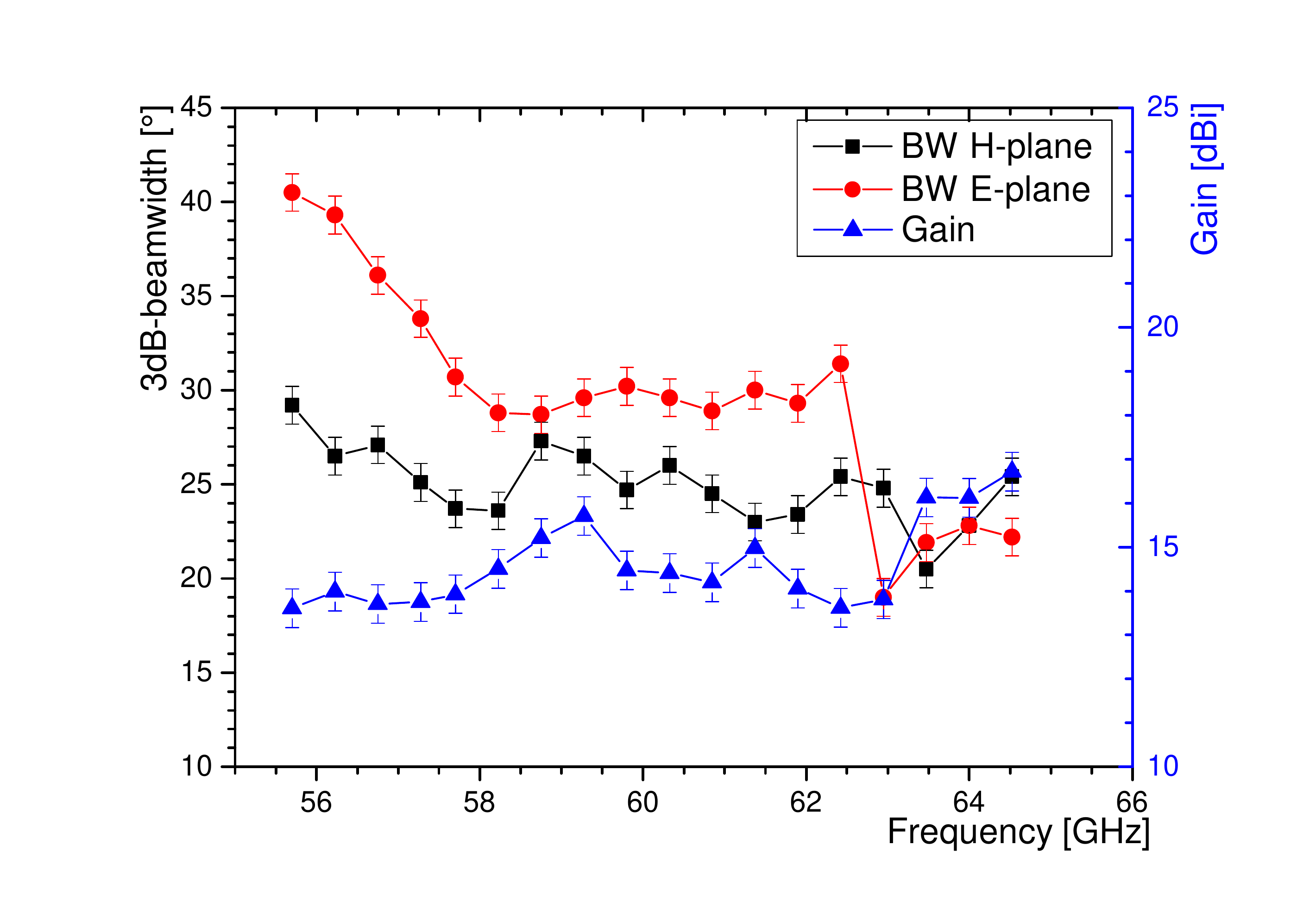}%antenna_kapton_gain_bw}
\caption{Measured beamwidth and gain of the Al-\kapton horn antenna, shown in Figure~\ref{fig:antenna_horn}.}
\label{fig:antenna_gain_bw}
\end{figure}

Figure~\ref{fig:antenna_direct} shows the gain of a \SI{3.5}{cm} long Al-\kapton horn
antenna in the H- and E-plane measured in the far field. 
The \SI{3}{dB}-beamwidth ${BW_{3dB}}$ is derived to be 
about \SI{25}{\degree} in the \linebreak H-plane and \SI{30}{\degree} in the E-plane. % at \SI{60}{GHz}.
A scan of the beamwidth and forward gain is performed in the frequency range from
\SIrange{55}{65}{GHz}, see Figure~\ref{fig:antenna_gain_bw}. 
In the whole range a gain $G \ge \SI{13}{dBi}$ is achieved. %, increasing with frequency. 
${BW_{3dB}}$ decreases accordingly in both planes as more intensity is focused in the forward direction. 
The errors on gain and beamwidth are derived from the following uncertainties: 
variations of the transmitted power over time, 
variations of signal attenuation along the coaxial cables due to bending, 
and geometrical uncertainties on the angle. 
%Above \SI{58}{GHz} a ${BW_{3dB}} \le \SI{30}{\degree}$ is achieved.
%RELATION BETWEEN FIELD BEAMWIDTHS AND GAIN UNDERSTOOD?
% Fouriertransformation: Betrag vom elektrischen Feld im Fernfeld proportional zu:
% sin(a/lambda *f(theta, phi))/(a/lambda *f(theta, phi)) * sin(b/lambda *f(theta, phi))/(b/lambda *g(theta, phi)) * h(theta,phi)
%http://www.wolframalpha.com/input/?i=sin%28x%29%2Fx%3B+1.2*sin%281.2*x%29%2F%281.2*x%29
%http://www.thefouriertransform.com/applications/radiation.php

The horn antennas presented here show high gain and high directivity.
They can be made of very little material such that they amount to less than
\SI{0.1}{\permil} of a radiation length.
A drawback is their volume making the mechanical integration into a
particle detector experiment potentially difficult. 
It should \linebreak further be noted that horn antennas require a primary feed antenna, for instance a single wideband patch or dipole antenna.

%Instead, they could be used to increase the gain of small but wideband single patches.

%\begin{figure}[h]
%\includegraphics[width=\columnwidth]{figures/antenna_eplane}
%\caption{Reflection loss.}
%\label{fig:antenna_e}
%\end{figure}
%
%\begin{figure}[h]
%\includegraphics[width=\columnwidth]{figures/antenna_hplane}
%\caption{Reflection loss.}
%\label{fig:antenna_h}
%\end{figure}
%
%\begin{figure}[h]
%\includegraphics[width=\columnwidth]{figures/antenna_gain}
%\caption{Reflection loss.}
%\label{fig:antenna_gain}
%\end{figure}

%Jakob und Letitia Durchführung 0
\subsection{Polarization}

%The second ansatz to minimize crosstalk is to exploit polarization.
Transmission tests with horn antennas with linearly \linebreak polarized waves are performed %(ARE ANTENNAS POLARIZING
%OR JUST KEEPING POLARIZATION?) 
and the polarization suppression factor $\mathrm{PSF}$ between orthogonal polarization states,
\begin{align}
\mathrm{PSF} &= \frac{|\vec{P}_{act}\times\vec{P}_{nom}|}{|\vec{P}_{act}\cdot\vec{P}_{nom}|}\,,%(\theta = \SI{90}{\degree})}{P(\theta = \SI{0}{\degree})}\,,
\end{align}
is measured. 
Here are $\vec{P}_{nom}$ the nominal polarization vector of the antenna and $\vec{P}_{act}$ the measured polarization vector. 
%The power $P$ is measured with the spectrum analyzer, the index denotes the polarization states of the two antennas to each other. % (parallel and orthogonal, correspondingly).  %polarization. %$\theta$ is the angle of rotation between the transmitting and receiving antenna. 
%$\theta = \SI{0}{\degree}$ corresponds to parallel polarization and $\theta = \SI{90}{\degree}$ to orthogonal polarization. 
%UNCLEAR; HOW IS DELTA I DEFINED?
%This value is an indication for the degree of polarization of the antennas.
%The measurement was performed by rotating the transmitting antenna by
%\SI{90}{\degree} with respect to the receiving antenna. % and measuring the received power.
The \linebreak distance between the antennas' apertures is about \SI{7.5}{cm}.
Two industrially produced brass horn antennas serve as reference. 
Results are shown in Table~\ref{tab:absorb_pol}.
%DELTA I=0 IF POLARIZATION PLANES IDENTICAL?

\begin{table}[tbp]
%		\footnotesize
		\centering
		\begin{tabular}{ccc}%c}%c}
		\toprule
		Horn antenna 1  & Horn antenna 2& $\mathrm{PSF}$  \\%& $X_0$ [m] \\%& $\SI{1}{cm}/X_0 [\%]$\\
		Transmitter  & Receiver & [dB]  \\%& $X_0$ [m] \\%& $\SI{1}{cm}/X_0 [\%]$\\
		\midrule
		Brass  & Brass  & \num[seperr]{-53.0(10)} \\%& \num[seperr]{6.1(1)} \\%& \num[seperr]{0.165(2)}\\
		\SI{4.5}{cm} Al-\kapton & Brass & \num[seperr]{-33.2(6)} \\%& \num[seperr]{8.3(2)} \\%& \num[seperr]{0.121(2)}\\
		\SI{3.5}{cm} Al-\kapton & Brass & \num[seperr]{-14.8(4)}  \\%& \num[seperr]{7.6(1)} \\%& \num[seperr]{0.138(1)}\\
		\SI{3.5}{cm} Al-\kapton & \SI{4.5}{cm} Al-\kapton  & \num[seperr]{-13.5(4)}\\% & \num[seperr]{8.8(1)} \\%& \num[seperr]{0.113(1)}\\
		\bottomrule
		\end{tabular}
		\caption{Polarization suppression factor ($\mathrm{PSF}$) of orthogonal polarization states using different antennas. Uncertainties are derived from measured intensity variations.} %Radiation length calculated from density measurement and $X_0 = \SI{44.77}{g/cm^2}$ for polyethylene.(, as there is no value for polyurethan.)}
		\label{tab:absorb_pol}
\end{table}

Compared to the brass antennas, 
the polarization suppression of the Al-\kapton horns % (USE CONSISTENTLY EITHER FOIL HORN OR AL-KAPTON) 
is smaller due to the shorter \linebreak length and mechanical imperfections.
Despite these \linebreak imperfections, pickup of signals from orthogonal polarization states is suppressed by more than \SI{10}{dB}. 
%However, even with these antenna prototypes, interfering signals could be significantly attenuated by more than \SI{10}{dB} by exploiting linear polarization.
%This is significantly lower most likely due to mechanical damage.
%hese tests show that interfering signals can be highly attenuated if one utilizes polarisation of antennas.
%For applications with linear polarisation this can be simply done by rotating the antennas for neighbouring links by \SI{90}{\degree}.

\subsection{Absorption and reflections}
\label{sec:absorb_foam}
%As a further method to reduce cross talk we discuss absorption of \SI{60}{GHz} waves. % and how reflections from a detector module can be attenuated.
Absorbing materials are in particular useful to avoid \linebreak transmission through
gaps in the detector and to attenuate \linebreak reflections from detector modules.
We tested different types and thicknesses of low density carbon impregnated foams, \linebreak which are commonly used as absorber for other microwave applications.
A list of the producer's material label (by ARC \linebreak Technologies), %\cite{url:arc})
the corresponding thickness and its density can be found in Table~\ref{tab:absorb_foam}. %and the calculated radiation length
All tested foams are of very low density and have large radiation lengths estimated to be \SIrange{6}{9}{m}.
For foam thicknesses of \SI{1}{cm} or less, the contribution to the material
budget of standard silicon detectors is rather small with $X/X_0 \lesssim 1\permil$.

\begin{table}[b]
%		\footnotesize
		\centering
		\begin{tabular}{lccc}%c}%c}
		\toprule
		Foam  & $d$ & $\rho$ &  $a$\\%& $X_0$ [m] \\%& $\SI{1}{cm}/X_0 [\%]$\\
		 & [mm] & [$\si{mg/cm^3}$] & [\si{dB/cm}] \\
		\midrule
		LS-11451-1 & \num{6.35}   & \num[seperr]{73.8(7)} & \num[seperr]{27.3(2)} \\%& \num[seperr]{6.1(1)} \\%& \num[seperr]{0.165(2)}\\
		LS-10122-9 \cite{arc:10122} & \num{12.70} & \num[seperr]{54.0(10)} & \num[seperr]{12.8(1)} \\%& \num[seperr]{8.3(2)} \\%& \num[seperr]{0.121(2)}\\
		LS-11297-1 & \num{19.05} & \num[seperr]{58.8(5)} & \num[seperr]{24.1(14)}\\%& \num[seperr]{7.6(1)} \\%& \num[seperr]{0.138(1)}\\
		LS-10640-1 \cite{arc:10640}& \num{25.40} & \num[seperr]{50.7(5)}& \num[seperr]{19.5(1)} \\% & \num[seperr]{8.8(1)} \\%& \num[seperr]{0.113(1)}\\
		\bottomrule
		\end{tabular}
		\caption[Properties of the tested graphite foams]{Properties
                  of the tested graphite foams: thickness $d$ (from ARC \mbox{Technologies}),
                  density $\rho$ and absorption loss $a$ at $f=\SI{60.7}{GHz}$ (measured).} %Radiation length calculated from density measurement and $X_0 = \SI{44.77}{g/cm^2}$ for polyethylene.(, as there is no value for polyurethan.)}
		\label{tab:absorb_foam}
\end{table}

The foam samples are placed between two antennas to \linebreak measure the transmission and reflection. 
The transmitted \linebreak intensity is normalized to the intensity measured in LOS configuration. % giving the transmission loss through the sample.
%As we are interested in measuring the attenuation of reflections from a detector module by applying graphite foam, we performed the reflection loss measurement with 

As the pore size of the foam is well below the signal wavelength of $\lambda \approx \SI{5}{mm}$, the surface of the foam can be treated as a plane surface. 
Transmission and reflection including polarization are described by the Fresnel equations \cite{Hecht}. 
%Thus, transmission and reflection losses can be described by the squared absolute value of the polarization dependent Fresnel equations \cite{Hecht} with a complex index of refraction 
The transmission loss $T$ through a layer of foam with a finite thickness $d$ is the product of the transmission losses at both surfaces and the absorption loss inside the foam described by the Beer-Lambert law:
\begin{align}
	T_{\bot / \parallel} &= \left| \,t_{\bot / \parallel}(\alpha, \beta, N_1, N_2) \right|^2 \cdot \left| \,t_{\bot / \parallel}(\beta, \alpha, N_2, N_1)  \right|^2 \cdot \mathrm{e}^{-a d}\,, \label{eq:T}
\end{align}
with the polarization dependent transmission loss  
\begin{align}
		t_{\bot}(\alpha, \beta, N_1, N_2)  &= \frac{2 N_1 \cos{\alpha}}{N_1 \cos{\alpha} + N_2 \cos{\beta}}\,,\\  											%\xrightarrow{\kappa = 0}	\frac{2 \sin{\beta}\cos{\alpha}}{\sin{\left(\alpha+\beta\right)}}\				
		t_{\parallel}(\alpha, \beta, N_1, N_2)  &= \frac{2 N_1 \cos{\alpha}}{N_2 \cos{\alpha} + N_1 \cos{\beta}}\,,  											%\xrightarrow{\kappa = 0}	\frac{2 \sin{\beta}\cos{\alpha}}{\sin{\left(\alpha+\beta\right)}\cos{\left(\alpha + \beta\right)}}						\label{eq:Elec_Fres2}
\end{align}
for waves polarized perpendicularly and parallel to the incident plane, respectively. 
The incident and the refracted angle are denoted by $\alpha$ and $\beta$ and the complex index of refraction $N_i$ is defined as 
%In Equation~\ref{eq:T}, the interchange of the variables $\alpha_1$ and $\beta_2$ corresponds to the interchange of the incident angle in air $\alpha$ and the refracted angle $\beta$ 
%and of the complex indices of refraction $N_i$, with 
\begin{align}
N_i &= n_i - i\kappa_i \,.
\end{align}
Here, $N_1 = 1$ for air and $N_2$ is the index of refraction of the graphite foam under test. 
The absorption coefficient $a$ is related to the imaginary part of the refractive index and the vacuum wavelength $\lambda_0 \approx \SI{5}{mm}$: 
\begin{align}
	a &= \frac{4 \pi \kappa}{\lambda_0}\,.
\end{align}	

\begin{figure}[tb]
\hspace{-1.5em}
\includegraphics[width=1.1\columnwidth]{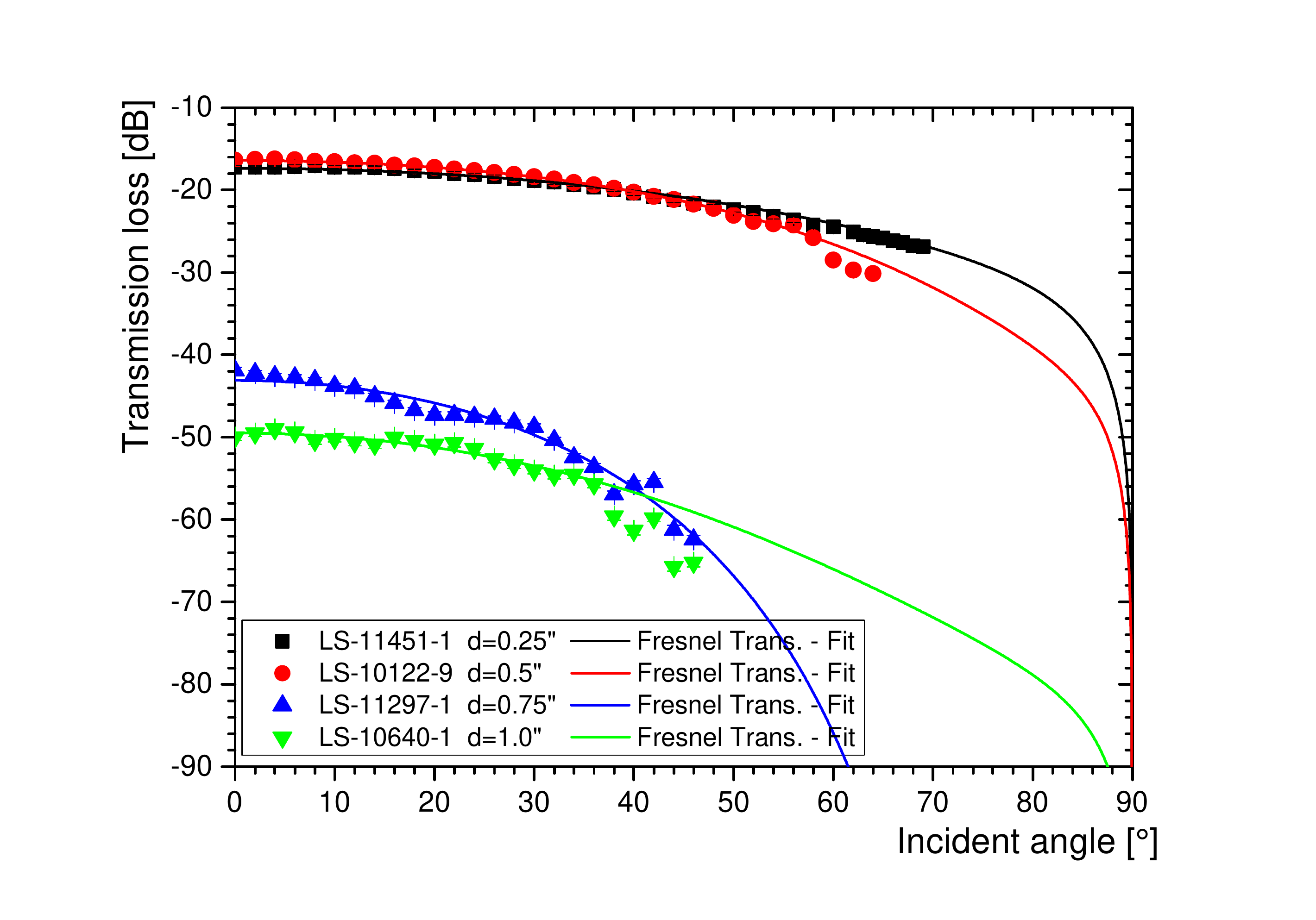}
\caption{Transmission loss of graphite foam samples at \mbox{$f = \SI{60.7}{GHz}$} for perpendicularly polarized waves. The uncertainties of $\num{0.2}-\SI{1.0}{dB}$ are due to intensity variations in the transmitted signal and the noise limited sensitivity of the spectrum analyzer. A fit of Equation~\ref{eq:T} is applied to all four data sets.}
\label{fig:absorb_trans}
\end{figure}

%DESRCRIPTION AND EXPLANTION OF FRESNEL FIT MISSING.
Figure~\ref{fig:absorb_trans} shows the measured transmission loss of the foam samples as function of the incident angle. 
All samples attenuate the transmitted signal by at least \SI[seperr]{15(1)}{dB} over the whole angular range. 
A $\chi^2$-fit of Equation~\ref{eq:T} is used to derive the absorption coefficient $a$.
%As follows from Equation~\ref{eq:T}, the absorption coefficient $\alpha$ can be deduced from the transmitted power.
Results are given in Table~\ref{tab:absorb_foam} in units of $\si{dB/cm}$. 
All samples show an absorption coefficient of \SI{20}{dB/cm} and higher, except foam LS-10122-9. 
Foam LS-11451-1 shows the highest absorption and therefore the best performance.

Similarly, absorption is measured over a larger frequency range in the \SI{60}{GHz}
band. No significant dependence on the frequency is found.
In addition, the homogeneity of the absorber materials is studied.  
A large piece of foam is used to study the fluctuations in the absorption
coefficient $a$ across the sample and variations up to \SI{3}{dB/cm} are measured.
Fluctuations could be due to inhomogeneities in the thickness, porosity and impregnation with graphite.

\begin{figure}[tb]
\hspace{-1.5em}
\includegraphics[width=1.1\columnwidth]{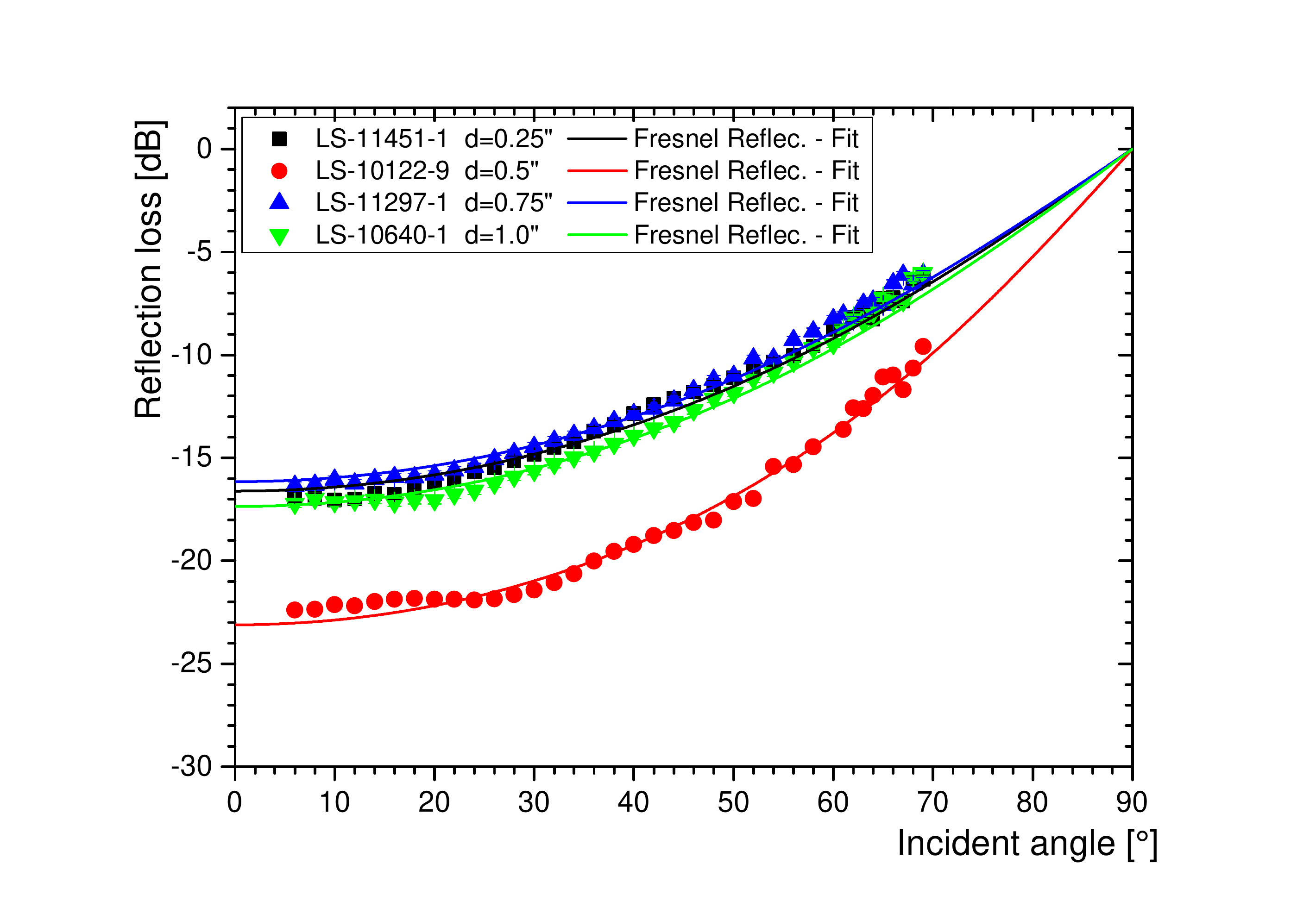}
\caption{Reflection loss of graphite foam samples at \mbox{$f =  \SI{60.7}{GHz}$} for perpendicular polarized waves. The uncertainties due to intensity variations are $\num{0.2}-\SI{0.3}{dB}$. A fit of Equation~\ref{eq:R} is applied to all four data sets.}
\label{fig:absorb_ref}
\end{figure}

Similar to the transmission, the reflection loss $R$ is described by %the corresponding Fresnel equations \cite{Hecht}: 
\begin{align}
	R_{\bot / \parallel} &= \left| r_{\bot / \parallel} \right|^2 \,, \label{eq:R}
\end{align}
with 
\begin{align}
		r_\bot &= \frac{N_1 \cos{\alpha} - N_2 \cos{\beta}}{N_1 \cos{\alpha} + N_2 \cos{\beta}}\,,\\		%\xrightarrow{\kappa = 0} 	- \frac{\sin{\left(\alpha-\beta\right)}}{\sin{\left(\alpha+\beta\right)}}					\label{eq:Elec_Fres1}\\
%\end{align}
%\begin{align}
		r_\parallel &= \frac{N_2 \cos{\alpha} - N_1 \cos{\beta}}{N_2 \cos{\alpha} + N_1 \cos{\beta}}\,, 	%\xrightarrow{\kappa = 0}		\frac{\tan{\left(\alpha - \beta\right)}}{\tan{\left(\alpha + \beta\right)}} \label{eq:Elec_FresBrewster}\\
\end{align}
for perpendicular and parallel polarization. 

The reflected intensity is measured and normalized to the power reflected by an aluminium plate. % which yields the reflection loss.
Figure~\ref{fig:absorb_ref} shows the reflection loss of the four foam 
samples at a frequency of \linebreak$f~=~\SI{60.7}{GHz}$ with waves polarized perpendicularly to the incident plane. 
The data is well fitted by the Fresnel ansatz. 
%A $\chi^2$-fit of Equation~\ref{eq:R} is applied to verify the Fresnel ansatz, which fits the data well.
All samples show a reflection loss below \SI{-10}{dB} up to large incident angles. 
The second thinnest sample, foam \mbox{LS-10122-9} (red curve), shows a significantly lower reflected intensity than the other samples. % and therefore the best performance. 
The measurement presented is repeated with the foam glued onto an aluminium plate and no difference is found. %No deviation from the measurement in air was found.
We conclude that, due to the high absorption coefficient of the foam, the total reflection loss is dominated by the first air - graphite foam transition. 
%Therefore, graphite foam is well suited to attenuate reflections off a detector module.

The methods discussed here %require no additional material in case of polarization, or almost no additional
%material for the other solutions, and 
are well suited to reduce reflections and cross talk
in tracking detectors without stressing the material budget.
For example, inadvertent transmission through gaps as observed for the endcap module in section~\ref{sec:trans} can be \linebreak easily damped by more than \SI{20}{dB} with a layer of graphite foam as thin as \SI{1}{cm}, corresponding to about \num{0.1}\,-\,\SI{0.2}{\%} of a radiation length. 
Reflections from detector modules and electronic \linebreak components can be attenuated up to large incident angles by \SI{10}{dB} and more by covering them with a layer of graphite foam. %, rather independent of the layer's thickness.

%To summarize this chapter, by exploiting highly directive antennas, polarization and 
%absorbers cross talk between \SI{60}{GHz} links can be largely reduced.
%Therefore, we are convinced that a wireless transceiver with antenna can be operated successfully when placed on a silicon detector module.

%\begin{figure}[h]
%\includegraphics[width=\columnwidth]{figures/xtalk_mat_foam_all_ref_perp}
%\caption{Reflection loss.}
%\label{fig:absorb_ref}
%\end{figure}
%
%\begin{figure}[h]
%\includegraphics[width=\columnwidth]{figures/xtalk_mat_foam_all_trans_perp}
%\caption{Transmission loss.}
%\label{fig:absorb_trans}
%\end{figure}

%\begin{figure}[h]
%\includegraphics[width=\columnwidth]{figures/xtalk_mat_foam_ins_fit_res}
%\caption{Frequency dependence.}
%\label{fig:absorb_freq}
%\end{figure}

%\blindtext
%\blindtext
%\blindtext
%\blindtext
%\blindtext
%\blindtext
%\blindtext
%\blindtext
%\blindtext

%\FloatBarrier

% !TeX root = ../main.tex
\section{Link density studies}\label{sec:links}

\begin{figure}[tb]
\centering
\includegraphics[width=\columnwidth]{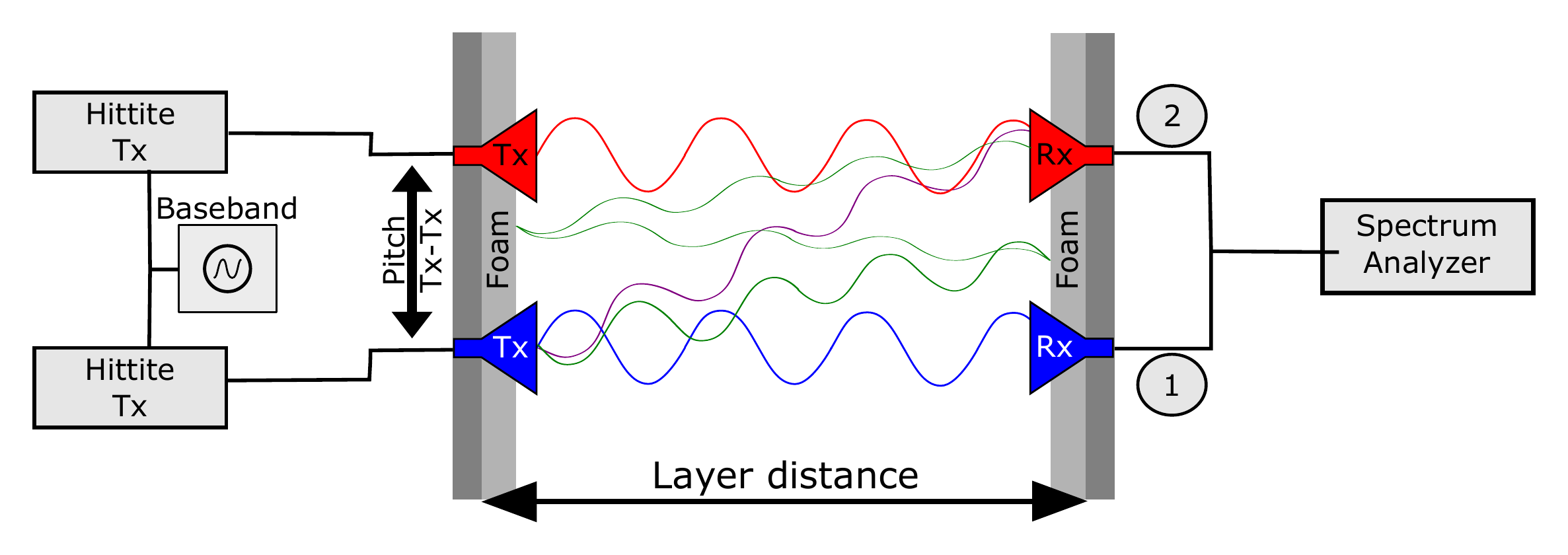}
\caption{Sketch of the setup to measure cross talk with two links between
  highly reflective aluminium layers. LOS coss talk is indicated as purple wave. }
\label{fig:links_foam_setup}
\end{figure}

By combining the different methods for reducing cross talk discussed in the
previous section, we study experimentally how close two parallel  \SI{60}{GHz}
links can be placed without disturbing each other, see Figure~\ref{fig:links_foam_setup}.  
%For this we measure the signal intensity emitted by two transmitters and received by two receivers and calculate the S/N.
%It is defined as the ratio of the signals 
We define as S/N the ratio of the signals from transmitter \#1 over transmitter \#2 
measured with the spectrum analyzer via the antenna of receiver \#1. %, see Figure~\ref{fig:links_foam_setup}. % sent from 
%The setup is illustrated in Figure~\ref{fig:links_foam_setup}.
%Furthermore, we performed bit error rate (BER) measurements.

We define the link quality to be sufficient %and good 
for an S/N of \SI{20}{dB} or higher.
This value corresponds to a theoretical bit error rate of \num{e-12}, or lower, for many modulation schemes, e.g.\ On-Off-Keying.
%This is motivated by the requirement On-Off-Keying to reach a bit error rate of \num{e-12} a S/N of \SI{17}{dB} is required \cite{seen in Hans Kristian paper!}.
%We add a \SI{3}{dB} safety margin for our measurements as we only take one neighbouring link into account, while it might be many more in the real application.

\subsection{Line of sight cross talk}

As LOS cross talk we define signals transmitted directly across different transmitter-receiver pairs. 
To study LOS cross talk, transmitters and receivers are placed in air without \linebreak reflecting layers behind the antennas. % in order to quantify the cross talk induced by LOS transmission only.  
The S/N is measured as function of the pitch, defined as the distance between the two transmitters. 
%The minimum pitch to achieve an S/N of \SI{20}{dB} between two links with identical linear polarization is studied. 
% by displacing the transmitting antenna perpendicular to the line of sight between the antennas.
The distance between transmitters and receivers is fixed to \SI{10}{cm} without antennas and both links have identical polarization. % with no horn antennas mounted.

\begin{figure}[tb]
\hspace{-1.5em}
\includegraphics[width=1.1\columnwidth]{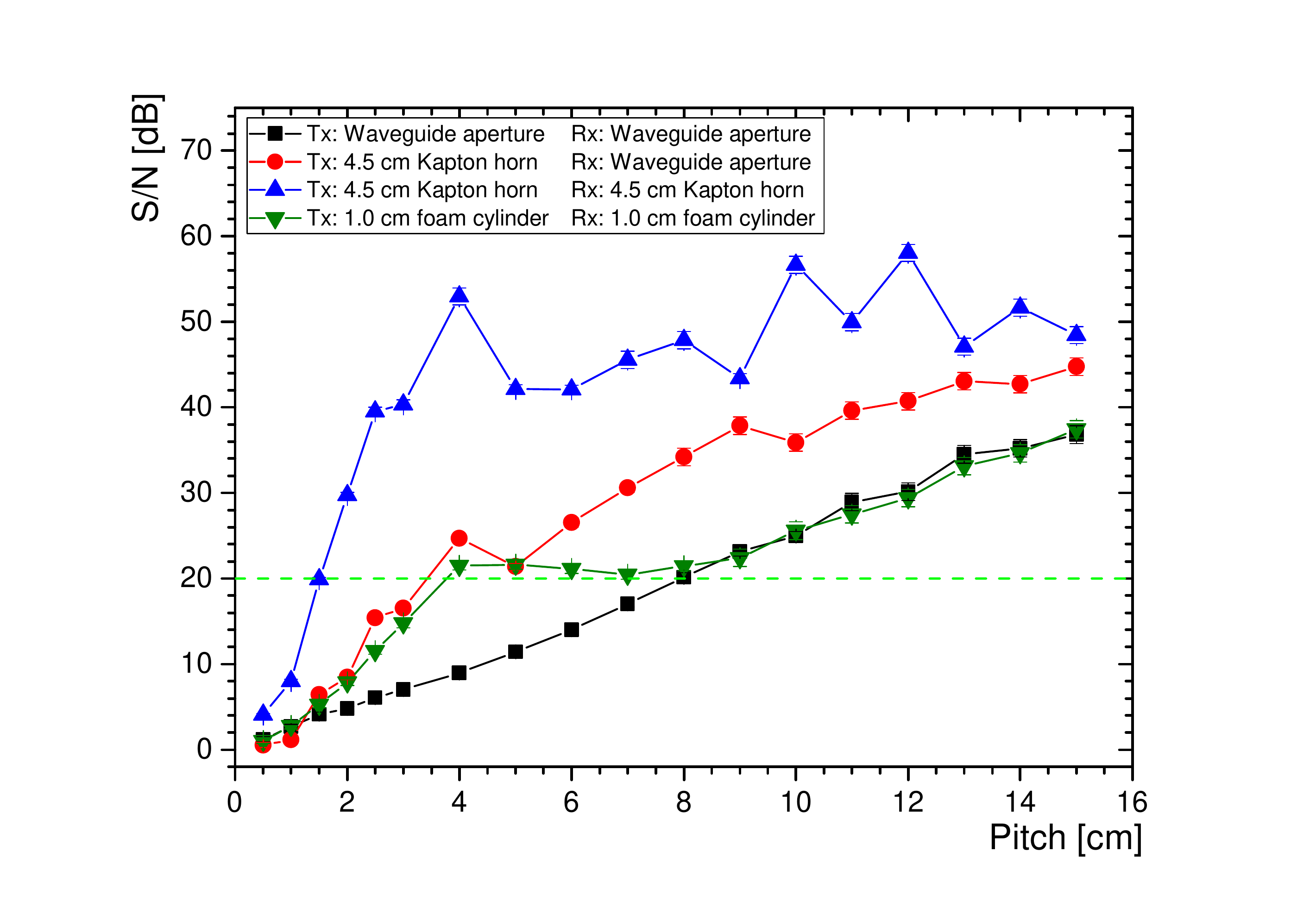}%snr_direct_parallel_H}
\caption{S/N in the radio frequency spectrum with LOS induced cross talk as function of the antenna pitch for different setups.}
\label{fig:links_direct_0}
\end{figure}

Figure~\ref{fig:links_direct_0} shows the S/N for different setups.
For reference, non-directive waveguide apertures are used ($BW_{3dB} \approx \SI{75}{\degree}$), for
which the S/N increases linearly as function of the pitch.
Without any measures a minimum pitch of about \SI{8}{cm} is \linebreak necessary to achieve an S/N of \SI{20}{dB}.
Using highly directive Al-\kapton horn antennas reduces the transmission distance in air to \SI{6.5}{cm} and \SI{3}{cm} for one or two antennas, respectively. 
With one antenna on the transmitting or receiving side, the minimum pitch is reduced to \SI{4}{cm}. % for a S/N $=\SI{20}{dB}$.
With \mbox{Al-\kapton} horn antennas installed on both sides, links can be placed as close as \SI{2}{cm} next to each other without significant interference effects.
Variations in the S/N with both horn antennas at pitches larger than \SI{4}{cm} are most likely due to the irregular radiation pattern in the side lobes (see Figure~\ref{fig:antenna_direct}).

\begin{figure}[tb]
\centering
\includegraphics[width=.4\columnwidth]{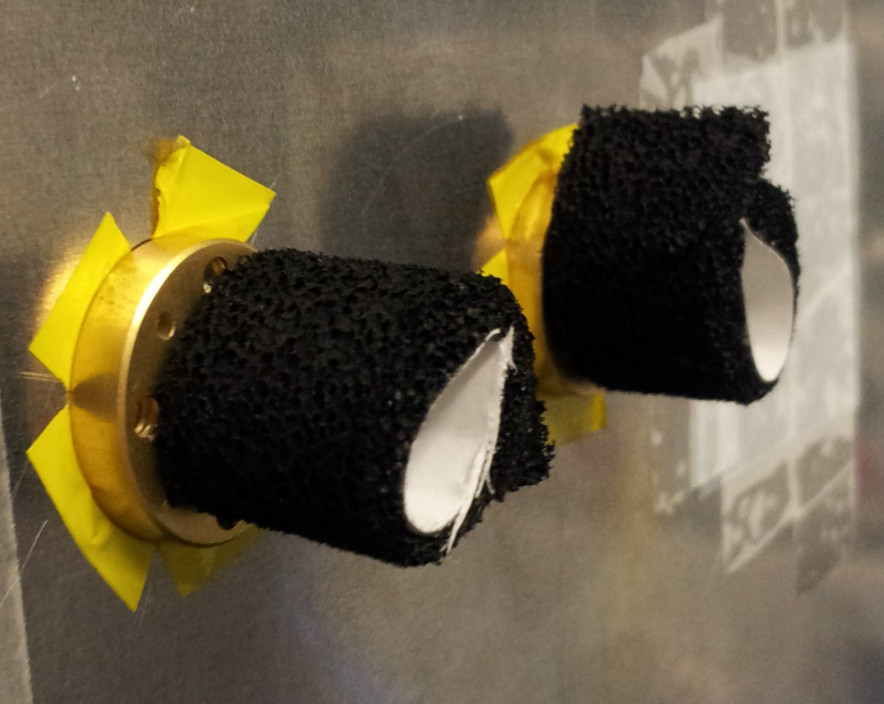}%snr_direct_parallel_H}
\caption{\SI{1}{cm} long hollow graphite foam cylinders shielding two wireless links. The paper covering the adhesive on the inside of the cylinder was tested to not affect the signal. }
\label{fig:links_foam_cylinders}
\end{figure}

%Even without using horn antennas cross talk can be reduced by absorber shields.
Without using any directive antennas cross talk can be \linebreak reduced just by absorber shields. 
For a test, all transmitters and receivers are equipped with \SI{1}{cm} long, hollow
graphite foam cylinders on top of the waveguide apertures to shield
lateral \linebreak radiation (see Figure~\ref{fig:links_foam_cylinders}).
It is found that shielded links can be placed as close as \SI{4}{cm}
next to each other to fulfill our S/N requirement. 
For small pitch values, the absorber shielding was found to be  as effective for cross talk reduction as horn antennas on one side.
For larger pitch, there seems to be no benefit from using shields. 
This could be due to rather strong side lobes generated by the cylinder's aperture.
%\color{red} WHY? \color{black} 
All results in Figure~\ref{fig:links_direct_0} are found by studying cross
talk in the H-plane. Similar results are obtained in the E-plane \cite{Dittmeier:Msc}.

\begin{figure}[tb]
\hspace{-1.5em}
\includegraphics[width=1.1\columnwidth]{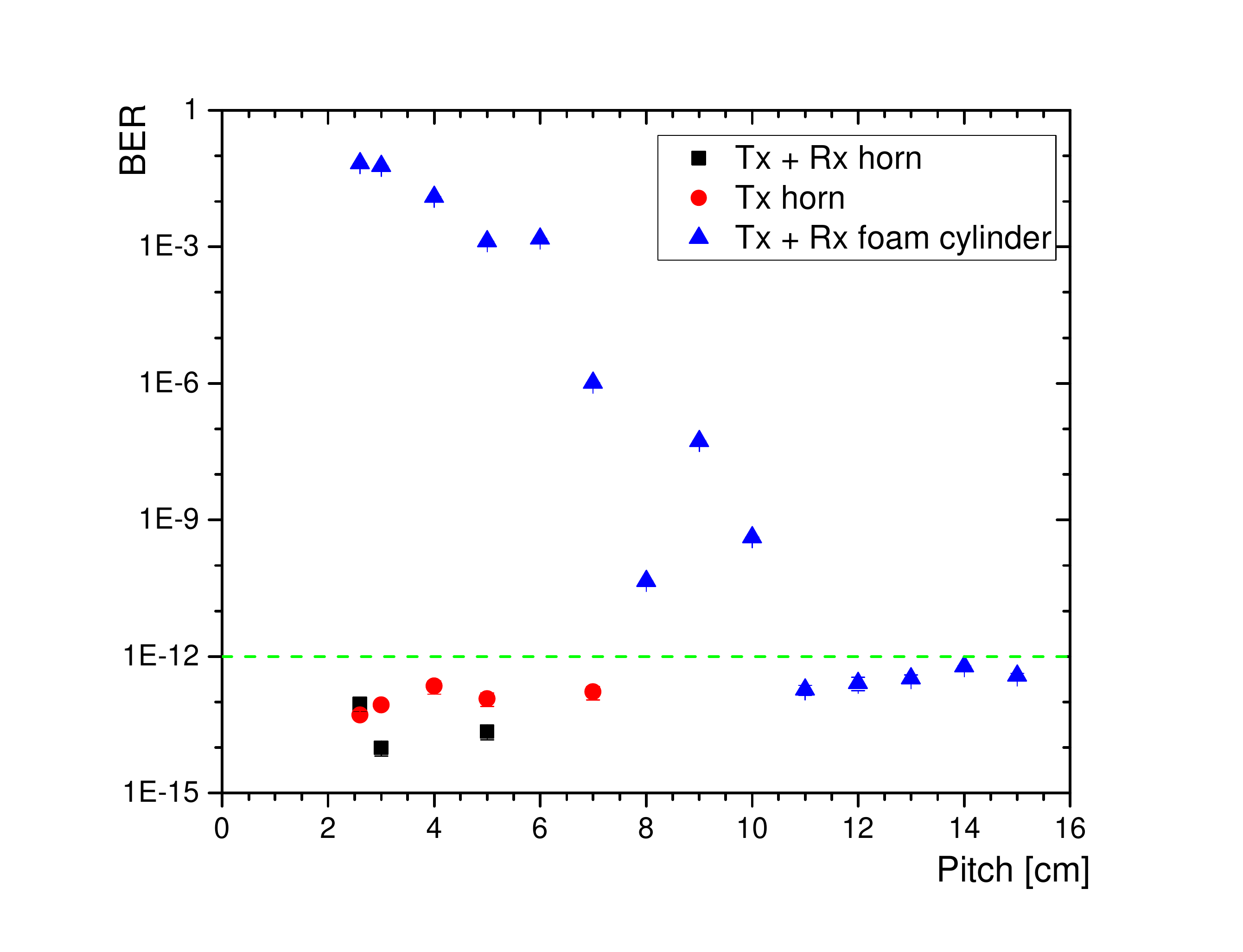}%data_cross_foam}
\caption{Influence of LOS cross talk on the bit error rate of a wireless data transmission, shown as function of the pitch between two parallel links. Distance between transmitter and receiver is set to \SI{10}{cm}. Both links are operated at the same carrier frequency.}
\label{fig:links_data_4}
\end{figure}

\subsection{Bit error rate tests}
For data transmission applications bit error rates (BER) give the relevant
figure of merit. % which are determined in the following for this setup.
Again, two links are placed at variable pitch with a fixed distance of \SI{10}{cm} between 
transmitter and receiver, not including the size of any antennas.
The BER is measured for a data rate 
of \SI{1.76}{Gb/s} %using the Hittite HMC6000/6001 transmitter and receiver
%chipset 
using Minimum Shift Keying (MSK) as modulation scheme.
Both links use the same polarization state and the same carrier frequency.
%Non-directive waveguide apertures are used as reference antennas.

BER measurements are performed with different configurations: horn antennas for transmitters only,
horn antennas for transmitters and receivers, and \SI{1}{cm} long
hollow graphite foam cylinders as shields instead of antennas.
%with the non-directive waveguide aperture in combination with \SI{1}{cm} long
%hollow gra-phite foam cylinders.
The BER \linebreak results are shown in Figure~\ref{fig:links_data_4}
as function of the pitch. 
With horn antennas bit error rates smaller than \num{e-12} are achieved 
for all studied pitches. 
Even with foam shields only, bit error rates below~\num{e-12} are achieved for a pitch larger than \SI{10}{cm}.	%with non-directive waveguide apertures 
Large variations below a pitch of \SI{10}{cm} are likely due to side lobes in the radiation pattern generated by the cylinder.

\subsection{Cross talk due to reflections}

Silicon detector modules are highly reflective as discussed in 
Section~\ref{sec:trans}. In a real tracking detector %non-absorbed 
\SI{60}{GHz} radiation is reflected between the silicon detector layers which leads to cross talk. 
By enclosing the volume with aluminium plates, the silicon detector environment is emulated, see Figure~\ref{fig:links_foam_setup}. 
%Therefore, above study is repeated with highly reflective aluminium plates
%emulating silicon detectors.
%As aluminium is fully reflective in the
%\SI{60}{GHz} range this setup serves as ``worst case'' scenario for reflections.
%As the SCT modules were measured to be highly reflective, reflections are included in the following study.
Two links are mounted through holes in each of the aluminium layers facing at a distance of \SI{10}{cm}. 
The pitch between the links is set to \SIlist{5;10;15}{cm}. 
Different configurations with directive horn antennas, graphite foam shields stuck 
onto the aluminium layers %(see Figure~\ref{fig:links_foam})
and orthogonal polarization states are tested. 
Again, waveguide apertures are used as non-directive reference antennas. 
The transmission distance in air is reduced to \SI{6.5}{cm} and \SI{3}{cm} when one or two horn antennas are applied, respectively. 
%A sketch of the setup is shown in Figure~\ref{fig:links_foam_setup}.

%\begin{figure}[h]
%\centering
%\includegraphics[width=.8\columnwidth]{figures/setup_foam}
%\caption{Two aluminium layers with transmitting and receiving waveguide apertures (as reference antennas) and graphite foam attached to attenuate reflections}
%\label{fig:links_foam}
%\end{figure}

\begin{figure}[tb]
\centering
\includegraphics[angle = -90, origin=c, width=\columnwidth]{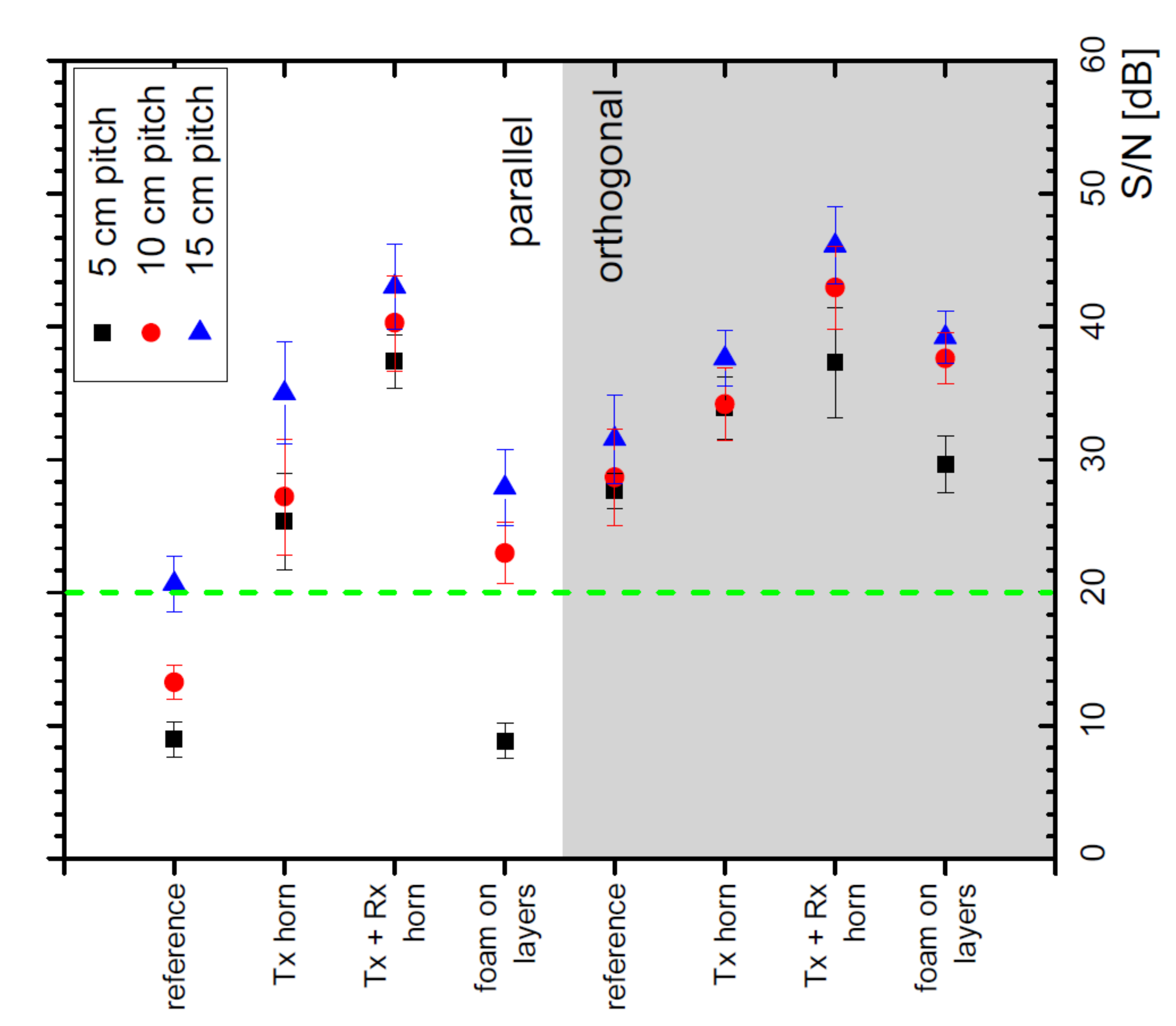} % snr_reflections_cut
\caption{S/N for two
  parallel links operated between two fully reflective aluminium layers at a
  distance of \SI{10}{cm}. 
Results are shown for parallel polarization states (top, white background) and
orthogonal polarization states (bottom, grey background).}
\label{fig:links_horn}
\end{figure}

Results of various measurements performed with this setup are shown in Figure~\ref{fig:links_horn}.
All antenna setups are measured with both polarization states (parallel and
orthogonal) between the two links.
We find that at a pitch of \SI{15}{cm} all studied setups give a satisfying S/N.
For a pitch of \SI{10}{cm} only the reference measurement with parallel polarization fails the S/N criterion.
%We find that without directive antennas (reference) and using parallel polarisation, a pitch of less than \SI{15}{cm} for parallel links does not seem feasible.
With directive antennas %for the transmitters (Tx) or for both transmitters and receivers (Tx + Rx), 
stable data transmission at a pitch of \SI{5}{cm} is possible for all studied configurations. 
%Without directive antennas, an S/N of more than \SI{20}{dB} can be achieved at a pitch of \SI{10}{cm} by applying graphite foam onto the reflective layers.
Using orthogonal polarization for neighboring links, a good S/N is 
obtained for all tested setups even without directive antennas 
%The largest improvement is made by using orthogonal polarisations for neighboring links. 
and a pitch of \SI{5}{cm} or less seems to be feasible. 
By combining directive antennas and absorbing foam the S/N ratio can be 
increased further. 
The operation of \SI{60}{GHz} links between highly reflective 
materials is possible even at a small pitch if directive antennas, absorber 
materials and/or polarization are exploited. 

\subsection{Frequency channeling}
\label{sec:links:ch}

The carrier frequency of the Hittite transmitter chip is tunable which allows
to use frequency channeling.
To measure quantitatively the BER as function of the carrier offset, 
two parallel links are positioned at a pitch of \SI{2.6}{cm} using the same
polarization state. 
Foam cylinders on top of the waveguide apertures are used for minimum shielding. 
The carrier frequency offset is varied in steps of \SI{500}{MHz}. % starting with
%the same carrier frequency for both links. 

\begin{figure}[tb]
\hspace{-1.5em}
\includegraphics[width=1.1\columnwidth]{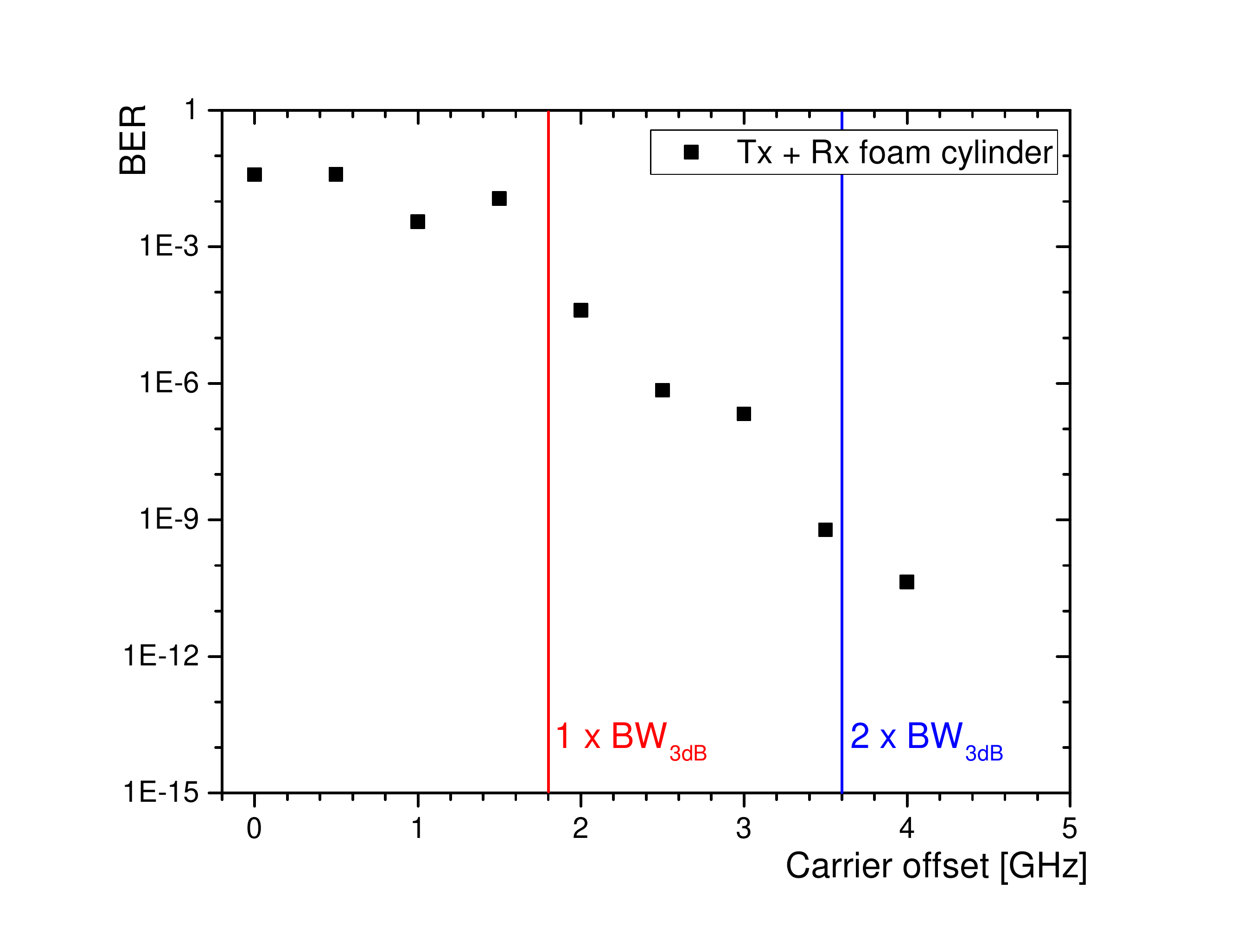}%data_cross_carrier}
\caption{Measured bit error rates of two parallel operated links at a pitch of \SI{2.6}{cm} as function of the carrier frequency offset. The links are operated with minimum shift keying at up to \SI{1.76}{Gbps}. No directive antennas are applied. The \SI{3}{dB}-bandwidth of the Hititte transceivers (\SI{1.8}{GHz}) is indicated as red line, twice the bandwidth as a blue line.}
\label{fig:links_data_0}
\end{figure}

The results are shown in Figure~\ref{fig:links_data_0}.
For frequency offsets smaller than the chipset's bandwidth 
($\pm\SI{1.8}{GHz}$) a plateau with a BER of a few percent is visible.
With increasing offset the BER decreases and reaches
\num{e-10} for an offset larger than twice the bandwidth. 
%for which the frequency bands don't overlap significantly anymore. 
With this setup, the usage of 3 frequency channels in parallel within the full \SI{60}{GHz} band seems possible. 
In order to actually benefit from channeling, filters with very sharp cutoff frequencies are required.

\subsection{Achievable link density}

From the results presented in this section, we can give an estimate on the possible link density. 
For this, we assume a tracking detector environment with distances of \SI{10}{cm} between layers and a minimum pitch of \SI{3.5}{cm} between links.  
%Without any measures, links can be placed safely at a pitch of \SI{15}{cm}.
%Applying horn antennas or in general directive antennas, we found that a pitch of \SI{5}{cm} is possible. 
%Exploiting polarization we gain another factor of $\sqrt{2}$ if we assume that links are placed equally distributed on a flat layer surface. 
%This gives us a minimum pitch of \SI{3.5}{cm}. 
%To estimate the data throughput, 
We first assume On-Off-Keying (OOK) as modulation scheme with a spectral efficiency of $\rho_{\mathrm{OOK}} \approx \SI{0.5}{b/s/Hz}$. 
%We further assume that we can 
Using the full \SI{9}{GHz} bandwidth per channel yields a data rate of \SI{4.5}{Gb/s} per link. 
The resulting data rate area density is about \SI{3.7}{Tb/(s \cdot m^2)}. 

%If one decides to use a bit more complex modulation scheme, e.g. Minimum Shift Keying (MSK), one can increase the spectral efficiency ($\rho_{MSK} = \SI{1}{b/s/Hz}$). 
A higher order modulation scheme can increase the spectral efficiency. 
MSK, providing $\rho_{\mathrm{MSK}} = \SI{1}{b/s/Hz}$, would be a viable candidate, as it can still be demodulated non-coherently. 
Using the same bandwidth we estimate that a data rate area density of about \SI{7.3}{Tb/(s \cdot m^2)} is feasible. %, which would be a challenge for other components interfacing to the wireless link.
%At this high bandwidth frequency channeling could be exploited. %But frequency channeling could be applied.

Exploiting frequency channeling, the link density can be further increased and we assume a minimum pitch of~\SI{2}{cm}. 
With an even higher order modulation scheme, three frequency channels with data rates of \SI{4.5}{Gb/s} per channel seem feasible. % if very sharp filters are used. 
%The channeling itself gives the possibility to increase the density % between the links again. 
%For three channels, we gain a factor of $\sqrt{3}$, which yields 
This results in a data rate area density of around~\SI{11}{Tb/(s \cdot m^2)}. 
However, frequency channeling and higher order modulation schemes typically come at the cost of increased complexity and power consumption. 
\section{Pickup of 60\,GHz noise}\label{sec:noise}

Deterioration of the detector performance due to pickup of noise from \SI{60}{GHz} communication is a potential worry for the operation of wireless links inside the detector.
However, cut-off frequencies of sensors and readout chips for silicon \linebreak detectors are typically below a few \si{GHz}. 
Therefore, no interference between the \SI{60}{GHz} links and tracking detector modules is expected. 
To demonstrate this, \SI{60}{GHz} irradiation tests are performed for different silicon strip and pixel detectors. 

\subsection{Test with silicon strip sensor prototypes}
%To proof the statement above we performed a \SI{60}{GHz} irradiation measurement using a test stand at the University of Freiburg which is described in \cite{Kuehn:Dimplon}.
%We tested if a wireless transmission in the \SI{60}{GHz} frequency band degrades the performance of silicon strip detector modules.
%In order to quantify this, a test was performed in cooperation with University of Freiburg.
Two ABCN endcap electronics hybrid prototypes for the phase-2 upgrade of the silicon tracking detector for the ATLAS experiment \cite{ATLAS:ABCN,Aliev2013210} are tested using a teststand at the University of Freiburg which is described in \cite{Kuehn:Dimplon}. 
One is a bare electronics hybrid, while the other one is connected to short silicon strip sensors with a strip length of \SI{2.4}{cm}. 
Each prototype comprises \num{12} fully functional ABCN readout chips~\cite{Kaplon:2008zz}. 

\begin{figure}[tbp]
\centering
\includegraphics[width=.9\columnwidth]{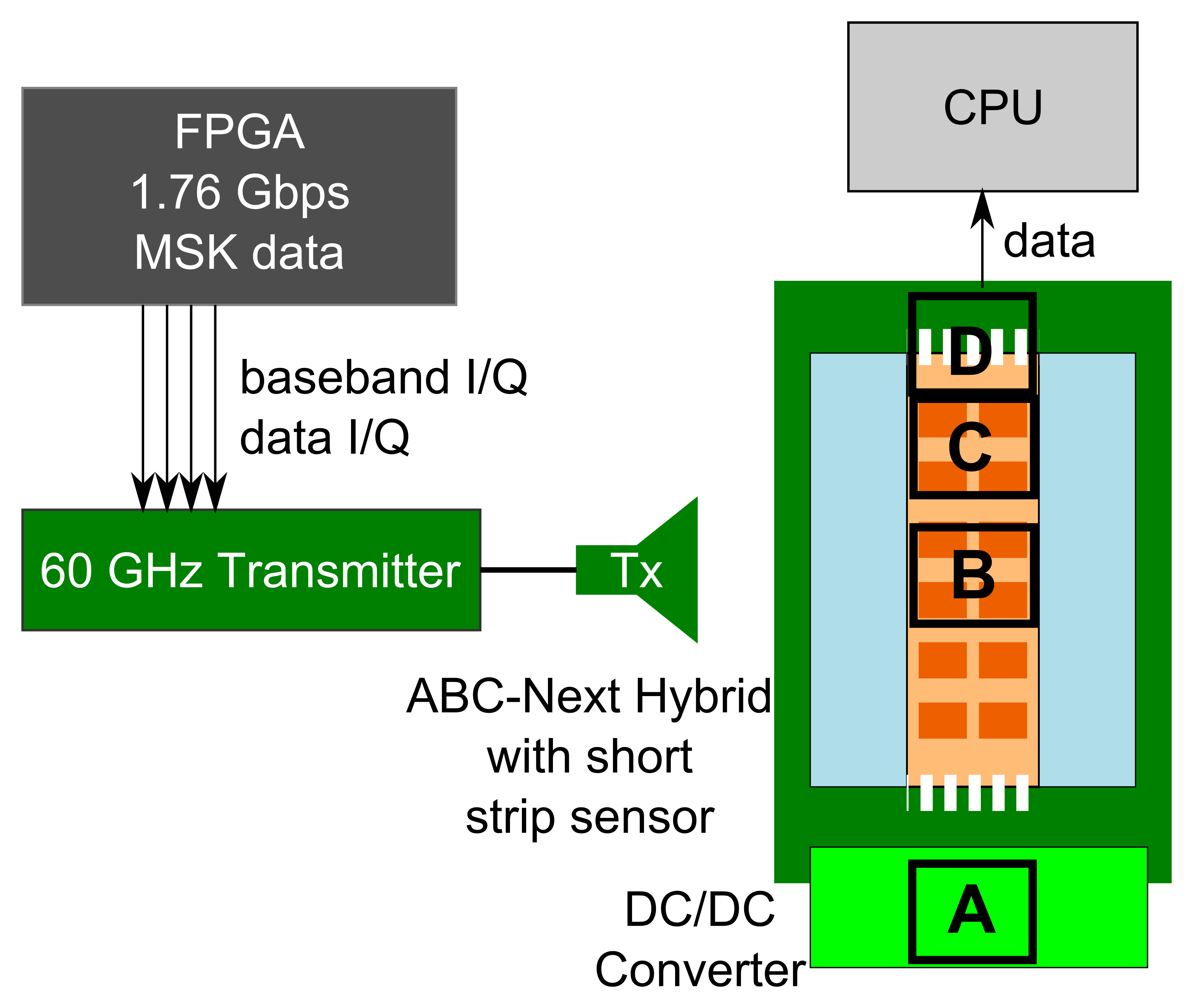}
\caption{Block diagram of the setup used for the irradiation test at Freiburg. A, B, C
  and D indicate different antenna positions used in the test. %, see also table~\ref{tab:noise_hybrid}.
The \num{12} readout chips are illustrated as brown boxes on the orange readout hybrid.}
\label{fig:noise_setup_strip_block}
\end{figure}

\begin{figure}[tb]
\centering
\includegraphics[width=.85\columnwidth]{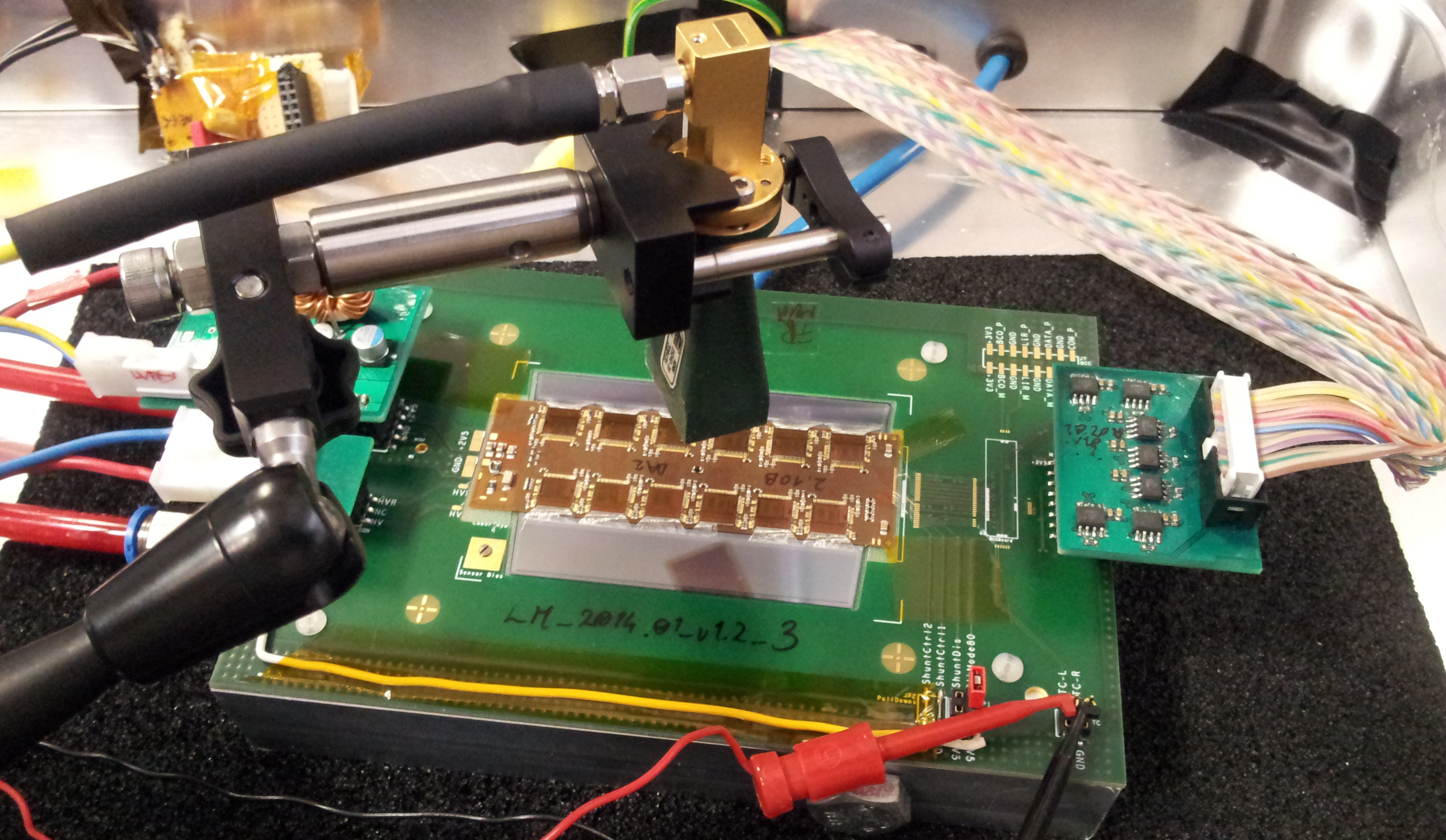}
\caption{A prototype for the ATLAS endcap tracking detector upgrade \cite{Aliev2013210} under irradiation of mm-waves.}
\label{fig:noise_setup_strip}
\end{figure}

The noise level of each readout channel is measured with a threshold scan using calibrated injected charges in units of equivalent noise charges (ENC).
Then, noise measurements are performed for each channel of the readout ASICs under wireless irradiation. 
Our \SI{60}{GHz} setup is used to irradiate the hybrid module using a \SI{20}{dBi} horn antenna from a distance of \SI{1}{cm}. 
%The \SI{60}{GHz} wireless signal was generated using a transmitter chip by Hittite (HMC6000) and modulated with a \SI{1.76}{Gb/s} pseudo random data stream using minimum shift keying as modulation scheme. 
%It was transmitted onto the hybrid with a \SI{20}{dBi} gain horn antenna placed at a distance of about \SI{1}{cm} above the module to create a typical, but rather high power signal for a short range communication.
%The peak intensity created by the transmitter was about \SI[seperr]{-15(3)}{dBm} at the carrier frequency. 
%The average intensity within a band of $\pm\SI{1}{GHz}$ around the carrier was about \SI[seperr]{-30(3)}{dBm}.
The \SI{60}{GHz} transmission power %integrated over the main lobe, which covers a frequency band of $\pm \SI{880}{MHz}$ around the carrier, was 
is about \mbox{$\SI[seperr]{-1.0(10)}{dBm}$}.
%to create a high intensity signal.
A block diagram and a photograph of the test setup are shown in Figure~\ref{fig:noise_setup_strip_block} and~\ref{fig:noise_setup_strip}, correspondingly. 
%As a reference, the noise measurements were performed without wireless irradiation beforehand and afterwards.
Four different positions are irradiated as indicated in Figure~\ref{fig:noise_setup_strip_block}: (A) the power converter, (B) and (C) the readout chips, and (D) the bonding wires. 
The irradiation measurements are performed with carrier frequencies of \SIlist{57;60;63}{GHz}. 
%Several measurements were performed focusing the wireless signal on different positions as indicated in Figure~\ref{fig:noise_setup_strip_block} and varying the wireless carrier frequency.

\begin{figure}[tb]
\centering
\includegraphics[width=\columnwidth]{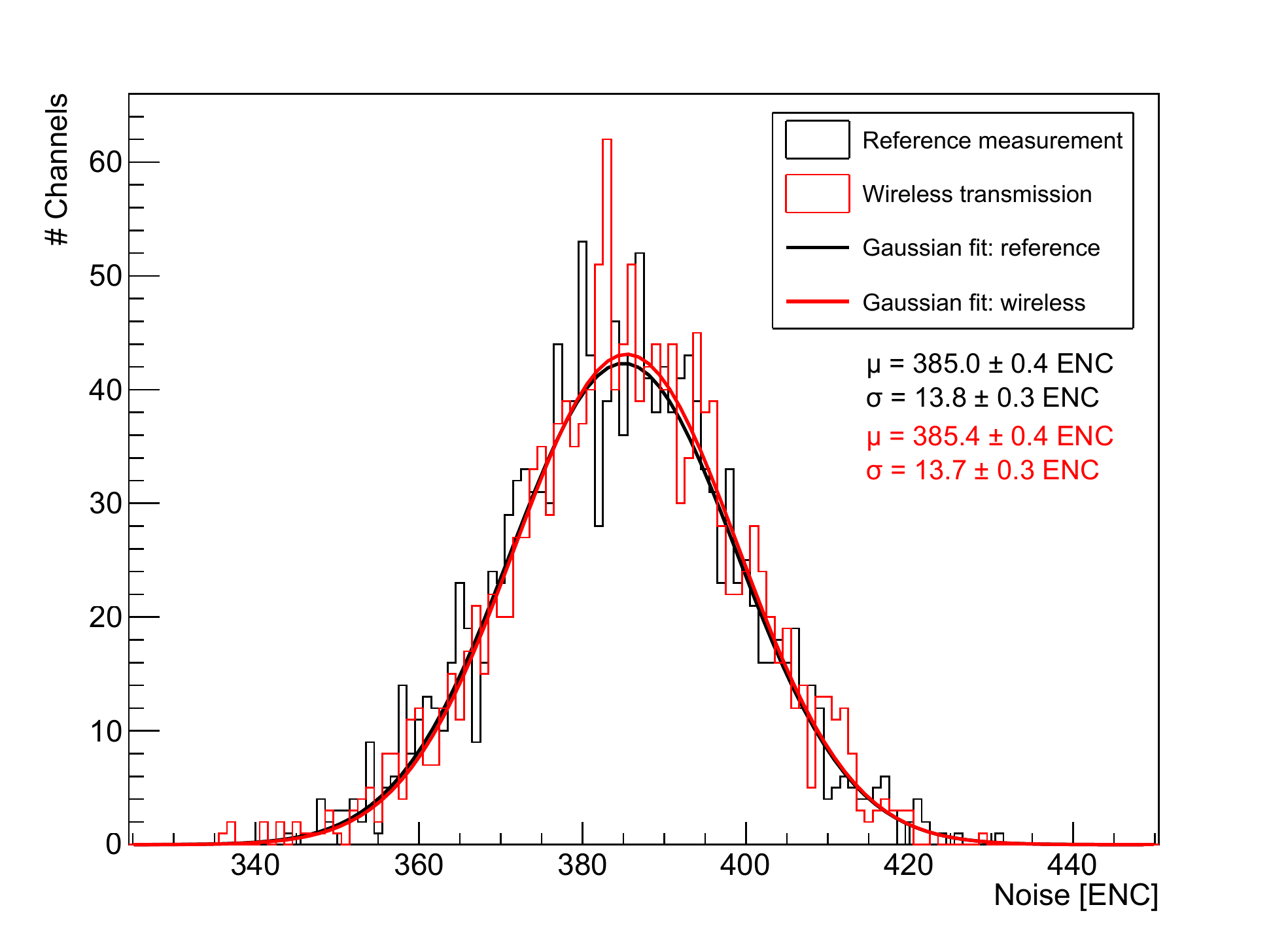}
\caption{Noise distribution of all channels of the \num{12} ABCN readout chips without strip sensors under irradiation. The reference measurement was performed without \SI{60}{GHz} irradiation. Mean $\mu$ and width $\sigma$ of the Gaussian fits are given. No significant difference is observed.}
\label{fig:noise_hybrid_dist}
\end{figure}

The noise level distribution of a reference measurement %(without \SI{60}{GHz} irradiation) 
and a measurement with \SI{60}{GHz} irradiation are shown in Figure~\ref{fig:noise_hybrid_dist} for the electronics hybrid without silicon strips. 
The noise distributions, see Figure~\ref{fig:noise_hybrid_dist}, are compatible with each other and
no significant differences are found in the mean and width of the distributions. 
Therefore, we set an upper limit of \SI{1}{ENC} at a \SI{95}{\%} confidence level on the increase of the average noise. 
Likewise, we do not observe any significant increase for the second prototype connected to silicon strips (average reference noise level: $\SI[seperr]{570(1)}{ENC}$; average noise with wireless irradiation: $\SI[seperr]{571(1)}{ENC}$). 
Thus, we set an upper limit of \SI{2}{ENC} at a \SI{95}{\%} confidence level on the average noise increase. % here. 
%Here, the noise was largely affected by a gradient in the ambient temperature over time as the box which contained the prototype had to be opened when the setup had to be adjusted. 

No significant rise in the noise level of the readout channels is observed at any carrier frequency or irradiated position. % of the antennas under test. 
As expected, no influence of the wireless signal on the noise level is found. 

% to be redone!
%A statistical analysis is performed to set a limit on the maximum noise level that could be induced by the wireless in a single readout channel without observing an increase at a \SI{95}{\%} confidence level using the $CL_s$ method \cite{0954-3899-28-10-313}.
%For the readout hybrid only (average background noise level: $\SI[seperr]{385(1)}{ENC}$) %$n_{H} = \SI{385}{ENC}$
% we can exclude any induced signal which is larger than $\SI[seperr]{43(1)}{ENC}$ at $\SI{95}{\%}\, CL_s$. %$n_{wireless, H} = \SI{43}{ENC}$ at $\SI{95}{\%}\, CL_s$.
%For the hybrid bonded to silicon strips, the observed average background noise was about $\SI[seperr]{570(1)}{ENC}$, we can exclude any influence larger than $\SI[seperr]{53(1)}{ENC}$ at $\SI{95}{\%}\, CL_s$.

%wireless: mean: 385.4 pm 0.4, sigma: 13.67 pm 0.3
%ref: mean:	385 pm 0.4, sigma 13.83 pm 0.3
%strips: ref: mean 569.7 pm 0.6, sigma 18.94 pm 0.5
%strips: wireless: mean 572.6 pm 0.6, sigma 19.08 pm 0.5
% avg: mean 571.4 pm 1.25
% kumulativer gauß: at 1.65 sigma
% symmetrischer gauß: at 2 sigma

\subsection{Test with HV-MAPS}

\begin{figure}[tb]
\centering
\includegraphics[width=\columnwidth]{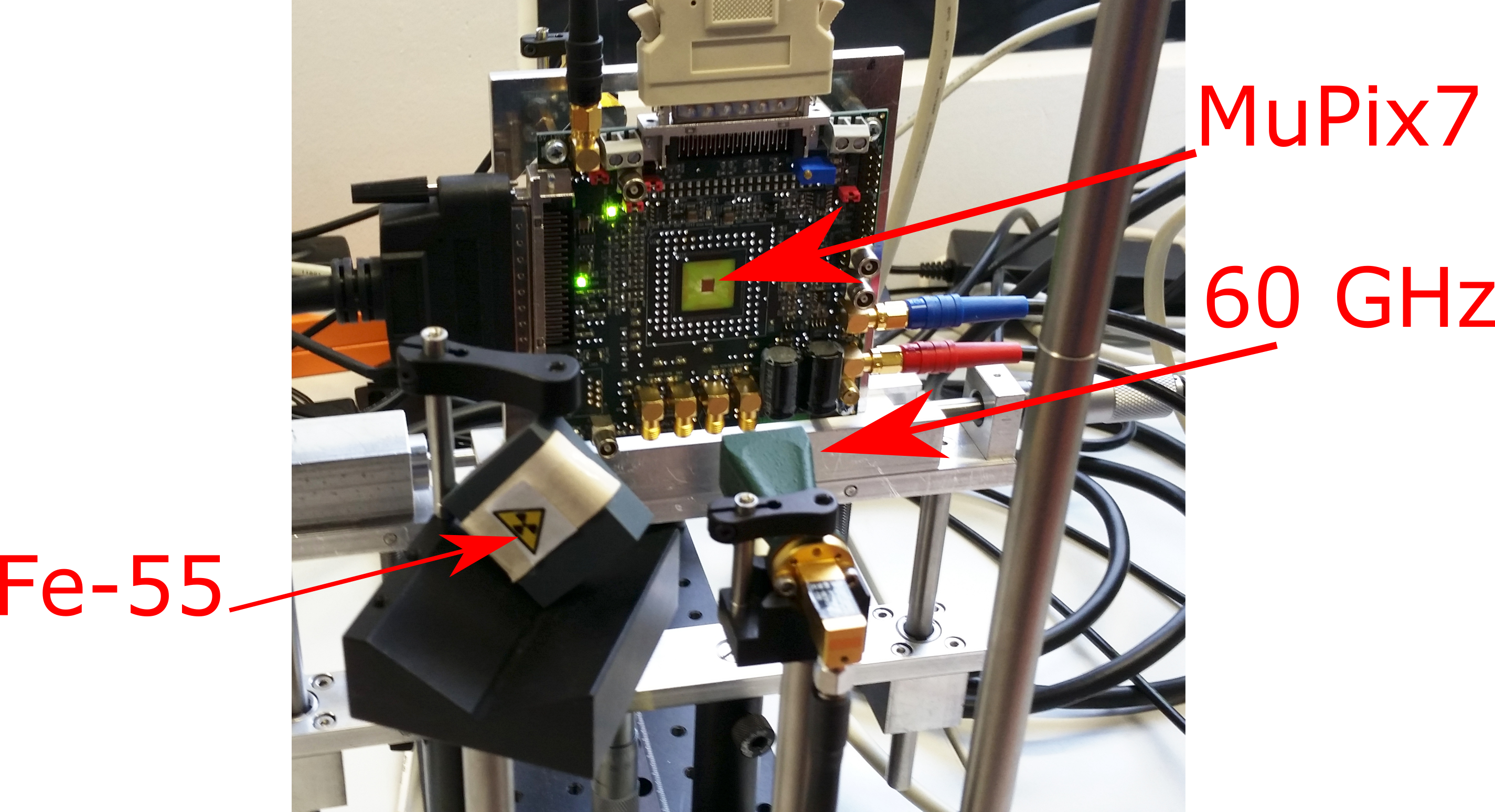}
\caption{The MuPix7 HV-MAPS prototype for the Mu3e experiment in the test setup irradiated with an $\mathrm{^{55}Fe}$ source. The wireless signal is transmitted using a horn antenna at a distance of about \SI{8}{cm} to the sensor.}
\label{fig:noise_mupix}
\end{figure}

\begin{figure}[tb]
\centering
\includegraphics[width=\columnwidth]{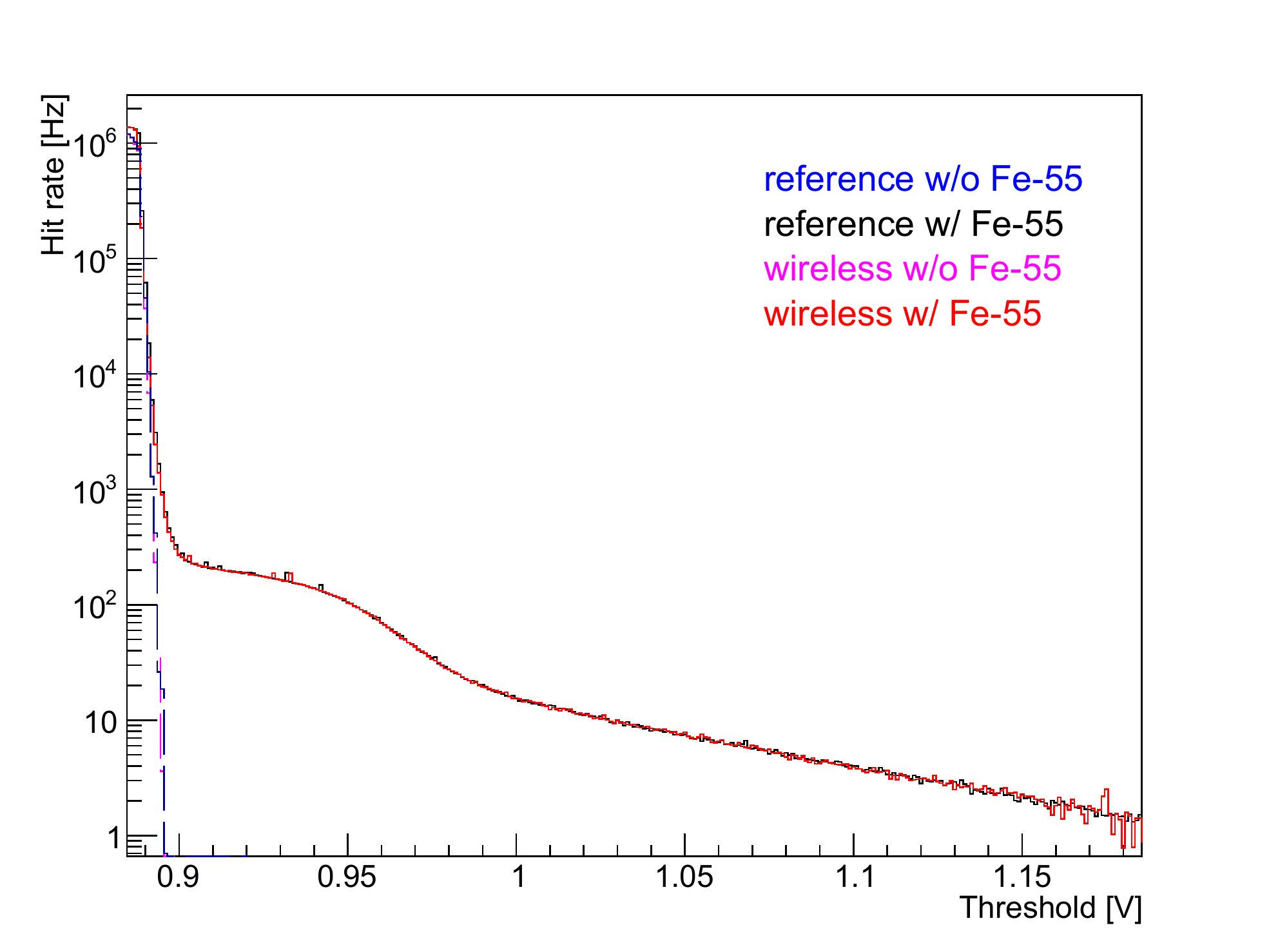}
\caption{Response of MuPix7 pixels to an iron $\mathrm{^{55}Fe}$ x-ray source measured with a threshold scan.}
%\caption{Relative response difference of the MuPix 7 to the iron source. Plotted is the ratio $\left(\# \mathrm{hits}_{on} - \# \mathrm{hits}_{off}\right)/ \# \mathrm{hits}_{off}$, where the index indicates if the wireless data transmission was turned on or off.}
\label{fig:noise_mupix_meas}
\end{figure}

We performed a similar measurement with a High Voltage-Monolithic Active Pixel Sensor (HV-MAPS) \cite{Peric:2007zz} prototype for the Mu3e experiment \cite{RP}, called MuPix \cite{Augustin:2015mqa,Augustin:2016hzx}. 
%The latest prototype, the MuPix7, is fully monolithic and includes a state machine and a fast serial readout operated at \SI{1.25}{Gb/s}. %is the first of its kind to incorporate its own internal state machine and a fast serial readout.
%The sensor features $\num{32}\times\num{40}$ pixels with a size of $\num{103}\times\SI{80}{\micro\metre^2}$ per pixel. 
%It is produced in a \SI{180}{nm} HV-CMOS process by AMS. 
%In the laboratory, we tested if the sensor is influenced by a wireless data transmission at \SI{60}{GHz}.
A picture of the \SI{60}{GHz} irradiation setup is shown in Figure~\ref{fig:noise_mupix}. 
%The \SI{60}{GHz} signal was generated using the Hittite HMC6000 chip transmitting a 8b10b encoded high definition video stream at a data rate of \SI{1.86}{Gb/s} with a transmitted power of about  $ P_{Tx} = \SI[seperr]{-1.0(10)}{dBm}$.
In order to see the influence of the wireless on the sensor operation, a threshold scan is performed with an $\mathrm{^{55}Fe}$ source, see Figure~\ref{fig:noise_mupix_meas}. 

%We measured the binary response of the sensor to a $\mathrm{^{55}Fe}$ source by means of the hit rate per pixel in dependence of the threshold applied in the comparator. 
%We did not observe any significant difference in the response to the iron source when the sensor was exposed to wireless signals in the \SI{60}{GHz} range  compared to a reference measurement, as can be seen in Figure~\ref{fig:noise_mupix_meas}.

Up to a threshold of about \SI{0.9}{\volt} the hit rate is dominated by noise. 
The S-curve response of the pixels to the iron source is clearly visible and has its turning point around \SI{0.96}{\volt} with a tail to higher signals. 
No effect of the \SI{60}{GHz} irradiation on the sensor performance is visible. 
%From a $\chi^2$-fit of an S-curve (error function) to both data sets we can deduce upper limits on the change of the distribution's turning point and width of \SI{1}{mV}. 
%Thus, no significant influence was found. 
%The results of a $\chi^2$-fit of an S-curve (error function) to both data sets are given in Table~\ref{tab:mupix_results}. 
%The parameters are in good agreement for both measurements. 
%No significant increase in the width $\sigma$ of the distribution or change in the position of the turning point $\mu$ could be observed. 
%
%\begin{table}[h]
%		\centering
%		\begin{tabular}{ccc}
%		\toprule
%		measurement  	&  $\mu [V]$  			& $\sigma [V]$ \\
%		\midrule
%		 reference	 	& \num[seperr]{0.9622(6)} 	& \num[seperr]{0.0273(2)}\\
%		 with wireless 	&\num[seperr]{0.9620(10)}   	& \num[seperr]{0.0267(2)}\\
%		\bottomrule
%		\end{tabular}
%		\caption{S-curve parameters of $\chi^2$-fits to the response to the source with and without wireless irradiation (reference). $\mu$ is the turning point of the S-curve; $\sigma$ is the width of the S-curve.}
%		\label{tab:mupix_results}
%\end{table}
Also, no influence on the power consumption of the sensor is found. % using a monitoring software for a HAMEG 4040 power supply. 
%From the observed current variations we can deduce an upper limit on the impact of the power consumption of \SI{1}{mW} for the whole test PCB which consumed about \SI[seperr]{750}{mW} in total. 
From these measurements, we can conclude that a wireless data transmission in the \SI{60}{GHz} band is safe with respect to the operation and performance of current silicon tracking detector sensors.

\section{Conclusion}\label{sec:conc}

The feasibility of a wireless readout system for a tracking detector was studied. 
Currently used ATLAS silicon strip modules are opaque to \SI{60}{GHz} radiation which allows to use the same carrier frequencies for communication between different, hermetically separated detector layers.
It was demonstrated, \linebreak using hybrid silicon strip modules from ATLAS and High \linebreak Voltage-Monolithic Active Pixel Sensors (HV-MAPS) from \linebreak Mu3e, that the performance of silicon detectors is not degraded under \SI{60}{GHz} irradiation.   
%It was demonstrated that the performance of a detector module prototype for the ATLAS upgrade is not degraded under irradiation of mm waves, as there was no additional noise observed. 
%The same was observed for a HV-MAPS prototype for the Mu3e experiment, the MuPix7.

By using directive antennas cross talk between neighboring links can be significantly reduced. 
Hence, high link densities are possible.
%Directive antennas are a key feature to achieve a high link density. 
%It was shown that cross talk between neighboring links can be reduced significantly with focusing antennas.
In addition, by exploiting linear polarization, links can be placed safely as close as \SIrange{2}{5}{cm} from each other for layer distances of \SI{10}{cm}.
Reflections and also transmission can be significantly attenuated using graphite foam, 
which would add only about \SI{1}{\permil} of a radiation length to the material budget due to the low density. 
%The amount of extra material by using foil horn antennas as presented as well as thin layers of graphite foam is of the order of \SI{0.1}{\permil} of radiation length.
% Thus, their contribution to the material budget would be almost negligible.
The contribution of antennas to the material budget is also very small if patch or horn antennas made of thin metalized foils are used. 
Frequency channeling %, i.e.\ using well separated frequency bands, 
is another option to decrease cross talk, but at the expense of a reduced bandwidth per link.
With all these measures combined we estimate that a data rate area density of \SI{11}{Tb/(s \cdot m^2)} is feasible for a \SI{60}{GHz} wireless readout system.
The study demonstrated that the \SI{60}{GHz} wireless technology is a very attractive alternative to wired and optical readout systems for future detector applications.

% !TeX root = ../main.tex
\section*{Acknowledgments}\label{sec:ack}

S.~Dittmeier acknowledges support by the \textit{International Max Planck Research School for Precision Tests of Fundamental Symmetries}. 
The authors would like to thank S.~Kühn and U.~Parzefall (University of Freiburg) for the opportunity to perform measurements with ATLAS SCT endcap modules.\linebreak
Furthermore, the authors would like to express their gratitude to R.~Brenner (University of Uppsala) for discussions and the supply of a spare ATLAS SCT barrel module.
%\bibliographystyle{elsarticle-num.bst}
%\section*{References}
\bibliographystyle{unsrt_collab_comma} %Style of Bibliography: plain / apalike / amsalpha / ...
%\printbibliography
\bibliography{main.bib}
\end{document}